\documentclass[%
reprint,
superscriptaddress,% this option gives proper header%
amsmath,amssymb,
aps,
pra,
floatfix,
]{revtex4-2}

\usepackage{graphicx}% Include figure files
\usepackage{dcolumn}% Align table columns on decimal point
\usepackage{bm}% bold math
\usepackage[%
pdftoolbar= true, %
pdfmenubar	=true, %	Anzeigen des Acrobat menu oder nicht
pdftitle	={On-demand storage and retrieval of single photons from a semiconductor quantum dot in a room-temperature atomic vapor memory}, %	Titel
pdfsubject	={Text}, %	Um was geht es
pdfauthor	={Maass, Barua}, %	Autor bzw. Autoren
pdfkeywords	={Text}, %	Stichwort1, Stichwort2 ...
pdfcreator	={Text}, %	Mit welcher Anwendung i.d.R. pdflatex
pdfproducer	={Text}, %	LaTeX with hyperref
colorlinks = true, % toggle textcolor links to colored border frame
linkcolor = black, % standard red
citecolor = black, % standard green
urlcolor = black, % standard magenta
    ]{hyperref} % add hypertext capabilities
\usepackage{comment}
\usepackage{xcolor} %use colored text %
\usepackage{lipsum}

\usepackage{tabularray}
\usepackage[T1]{fontenc} % Hinzugefügt von Benni
\usepackage{float} % Hinzugefügt von Benni
\graphicspath{{./Figures/}} % Hinzugefügt von Benni
\usepackage[caption=false]{subfig}

\begin{document}

\preprint{APS/123-QED} %seems to toggle style and footnotes in header %

%\title{Room-temperature ladder-type optical memory compatible with single photons from %InGaAs 
%quantum dots}% 
\title{On-demand storage and retrieval of single photons from a semiconductor quantum dot in a room-temperature atomic vapor memory}% 

\author{Benjamin Maa\ss}
 \altaffiliation[The authors contributed equally to this work.]{}%
 \affiliation{Institute of Optical Sensor Systems, German Aerospace Center (DLR), Rutherfordstra{\ss}e 2, 12489 Berlin, Germany}%
 \affiliation{Institute of Solid State Physics (IFKP), Technische Universit\"at Berlin, Hardenbergstra{\ss}e 36, 10623 Berlin, Germany}
    %\altaffiliation[Also at ]{Physics Department, XYZ University.}%
 \author{Avijit Barua}
 \altaffiliation[The authors contributed equally to this work.]{}%
 \affiliation{Institute of Solid State Physics (IFKP), Technische Universit\"at Berlin, Hardenbergstra{\ss}e 36, 10623 Berlin, Germany}
 \author{Norman Vincenz Ewald}%
 \affiliation{Institute of Optical Sensor Systems, German Aerospace Center (DLR), Rutherfordstra{\ss}e 2, 12489 Berlin, Germany}%
 \affiliation{Physikalisch-Technische Bundesanstalt (PTB), Berlin, Germany}
 \author{Elizabeth Robertson}
 \affiliation{Institute of Optical Sensor Systems, German Aerospace Center (DLR), Rutherfordstra{\ss}e 2, 12489 Berlin, Germany}%
 \author{Kartik Gaur}
 \affiliation{Institute of Solid State Physics (IFKP), Technische Universit\"at Berlin,  Hardenbergstra{\ss}e 36, 10623 Berlin, Germany}
   \author{Suk In Park}
 \affiliation{Korea Institute of Science and Technology (KIST), Seoul, Republic of Korea}
  \author{Sven Rodt}
 \affiliation{Institute of Solid State Physics (IFKP), Technische Universit\"at Berlin,  Hardenbergstra{\ss}e 36, 10623 Berlin, Germany}
   \author{Jin-Dong Song}
 \affiliation{Korea Institute of Science and Technology (KIST), Seoul, Republic of Korea}
  \author{Stephan Reitzenstein}
  \email{stephan.reitzenstein@physik.tu-berlin.de}
 \affiliation{Institute of Solid State Physics (IFKP), Technische Universit\"at Berlin,  Hardenbergstra{\ss}e 36, 10623 Berlin, Germany}
 \author{Janik Wolters}
   \email{janik.wolters@dlr.de}
 \affiliation{Institute of Optical Sensor Systems, German Aerospace Center (DLR), Rutherfordstra{\ss}e 2, 12489 Berlin, Germany}%
 \affiliation{Physikalisch-Technische Bundesanstalt (PTB), Berlin, Germany}
 \affiliation{Institute of Optics and Atomic Physics (IOAP), Technische Universit\"at Berlin, Hardenbergstra{\ss}e 36, 10623 Berlin, Germany}

%\collaboration{HQSys DLR TU Berlin}%\noaffiliation

\date{\today}% It is always \today, today,

\begin{abstract}
Interfacing light from solid-state single-photon sources with scalable and robust room-temperature quantum memories has been a long-standing challenge in photonic quantum information technologies due to inherent noise processes and time-scale mismatches between the operating conditions of solid-state and atomic systems. Here, we demonstrate on-demand storage and retrieval of single photons from a semiconductor quantum dot device in a room-temperature atomic vapor memory. A deterministically fabricated InGaAs quantum dot light source emits single photons at the wavelength of the cesium D1 line at 895\,nm which exhibit an inhomogeneously broadened linewidth of 5.1(7)\,GHz and are subsequently stored in a low-noise ladder-type cesium vapor memory. We show control over the interaction between the single photons and the atomic vapor, allowing for variable retrieval times of up to 19.8(3)\,ns at an internal efficiency of $\eta_\mathrm{int}=0.6(1)\%$. Our results significantly expand the application space of both room-temperature vapor memories and semiconductor quantum dots in future quantum network architectures.		
\end{abstract}

%\keywords{quantum memory, quantum dot, single photon, quantum communication, photon storage, broadband, atomic vapor, room temperature}%Use showkeys class option if keyword

\maketitle

\section{Introduction}
	High-fidelity quantum networks require an architecture that connects individual quantum nodes via photonic quantum channels \cite{Kimble.2008}. 
	In this context, quantum memories play a pivotal role in distributing entanglement over large distances\cite{Duan.2001}, buffering local network nodes and increasing the networks's throughput. Due to their scalability and robustness, room-temperature atomic vapor memories are a particularly well-suited memory platform \cite{Mehdi.2022,Jutisz.2024}. Semiconductor quantum dots (QD), on the other hand, have emerged as a highly promising platform for the deterministic generation of single photons and entangled photon states of light in the field of photonic quantum information processing \cite{Heindel2023}. In particular, InGaAs QDs operating in the near-infrared (NIR) spectral range are highly suitable for integration into quantum networks due to their ability to emit single photons with high purity and indistinguishability \cite{Pascalesenellart,Yu2023}. Controlled interactions between such heterogeneous systems can be the backbone of future quantum network applications such as quantum-secure communication \cite{DanielAQTReview,HKLo_nature20214,Trotta2021}, remote sensing\cite{Komar.2014,Zhao.2021, Nichol.2022} or distributed quantum computing 
 \cite{Wehner2018DistributedQC}.
	
	Interfacing QD-based single-photon sources (SPS) with atomic vapor-based quantum memories presents substantial challenges due to bandwidth mismatches between solid-state and atomic systems, intrinsic noise processes and the need for matching of operational wavelengths with sub-GHz accuracy \cite{Warburton2020}. 	
	Nevertheless, significant progress has been achieved in this field. This includes the wavelength-matched interaction and delay of QD single photons in alkali vapor \cite{Akopian.2011, Wildmann.2015, Kroh.2019, Bremer:2020}, delay of two-photon Fock states \cite{Vural.2021} and delay of entangled photons \cite{Trotta.2016}. However, these demonstrations only delayed the single photons but did not allow for on-demand control of the interaction with the alkali vapor. The latter requires a suitable low-noise and bandwidth-matched optical memory.

        \begin{figure*}
        \includegraphics[width=1.8\columnwidth]{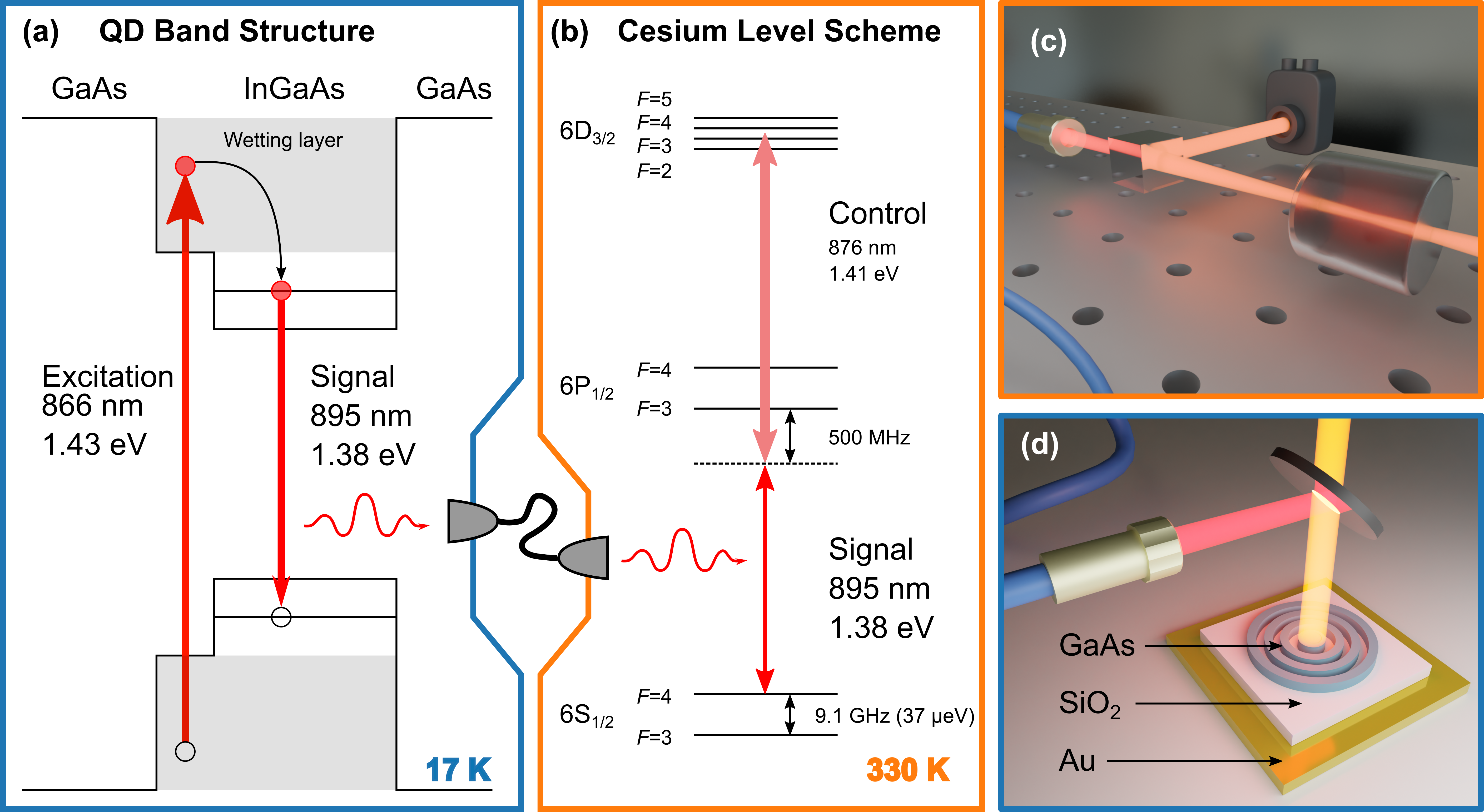}
	\caption{Interconnection of QD SPS with atomic quantum memory. (a)~Electronic band structure of the InGaAs QD; (b)~Atomic level scheme of cesium including the D1 signal and control transitions; Artistic impression of (c) the quantum memory and (d) the QD-hCBG SPS.}
	\label{fig:CBG}
    \end{figure*}

 Previous works on ladder-type memories in room-temperature vapors \cite{Kaczmarek.2018, RanFinkelstein.2018} have shown simultaneous ultra-low noise and high-bandwidth operation which is a key requirement for storing QD photons with large linewidth in optical memories. Indeed, several alkali vapor memories have been experimentally identified to be readily compatible with single photons emitted from QD devices \cite{Wolters.2017, Thomas.2023, Maaß.2024}. Most recently, telecom single photons from an InAs QD have been stored in a hot atomic vapor memory in a proof-of-concept experiment\cite{Thomas.2024}. However, the level configuration of the used memory limited the effectively non-variable storage time to 800\,ps, i.e. on the timescale of the exciton lifetime. For synchronization and buffering a time-bandwidth product $B\gg1$ and real time controllability of the on-demand memory read-out are required. A QD-memory interface that provides such utility remains a milestone yet to be achieved.  
 
	In this work, we present such an interface between an InGaAs QD device and a room-temperature Cs vapor memory, showing on-demand storage and retrieval. The QD is deterministically integrated \cite{Rodt_2021,Rodt_2020} into a hybrid circular Bragg grating (hCBG) cavity\cite{Rickert:19,YaoBCBG} and wavelength-aligned with the Cs D1 line at 895\,nm. The ladder-type memory allows for programmable storage times of up to $\tau_\mathrm{s}=19.8(3)$\,ns, exceeding the spontaneous emission lifetime of the QD $\tau_\mathrm{QD}=1.4(1)$\,ns by more than one order of magnitude, resulting in a time-bandwidth product of $B=\tau_\mathrm{s}/\tau_\mathrm{QD}= 14(1)$. The QD photons are stored and retrieved with an internal efficiency of $\eta_\mathrm{int}=0.6(1)\%$.
 
  \section{Experimental setup}
  
	This experiment consists of two heterogeneous parts: the cryogenically cooled QD-hCBG SPS and the room-temperature atomic vapor memory, as depicted in Fig. \ref{fig:CBG}. Initially, an optical setup collects the single photons emitted by the QD device and couples them into a single-mode fiber to be connected to the quantum memory setup. In the following section, we first describe the fabrication of the QD device and its optical properties before presenting the QD-memory interface.

	\subsection{QD single-photon source}
 \label{sec:QD_source}
The QD-hCBG SPS employed in this study is based on a sample containing InGaAs QDs, grown by molecular beam epitaxy (MBE)\cite{MBEgrowth}. By employing a flip-chip wafer bonding process, a thin membrane of GaAs with the embedded QD-layer is hybridly integrated with silicon dioxide ($\text{SiO}_2$) as a dielectric layer and a backside gold mirror as a reflector to redirect the downward propagating light towards the top \cite{Wang2019,Jinliu2019CBG}. To further improve the collection efficiency and to generate a near-Gaussian far-field emission profile of the source, the QD sample is integrated with a CBG cavity which comprises a central mesa surrounded by three periodic concentric rings\cite{Jinliu2019CBG,YaoBCBG,Nawrath_Michler, Shih:23}.
The fabrication of the device is based on the \textit{in-situ} electron beam lithography (iEBL) technique which allows for straightforward deterministic integration of QDs into the resonator structure\cite{Rodt_2021,donges2024}. The design parameters for the hCBG cavity were numerically optimized using a finite element method (FEM)\cite{Burger_JCM,JCMWebsite}.

\begin{figure*}
    \subfloat{
    \includegraphics[width=0.9\columnwidth]{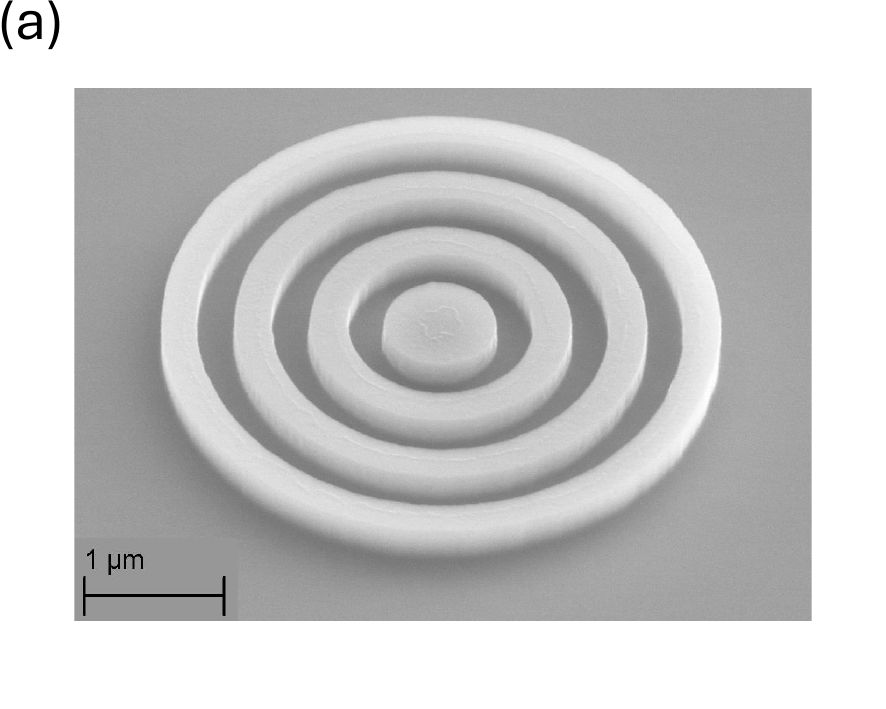}}
    \subfloat{\includegraphics[width=0.9\columnwidth]{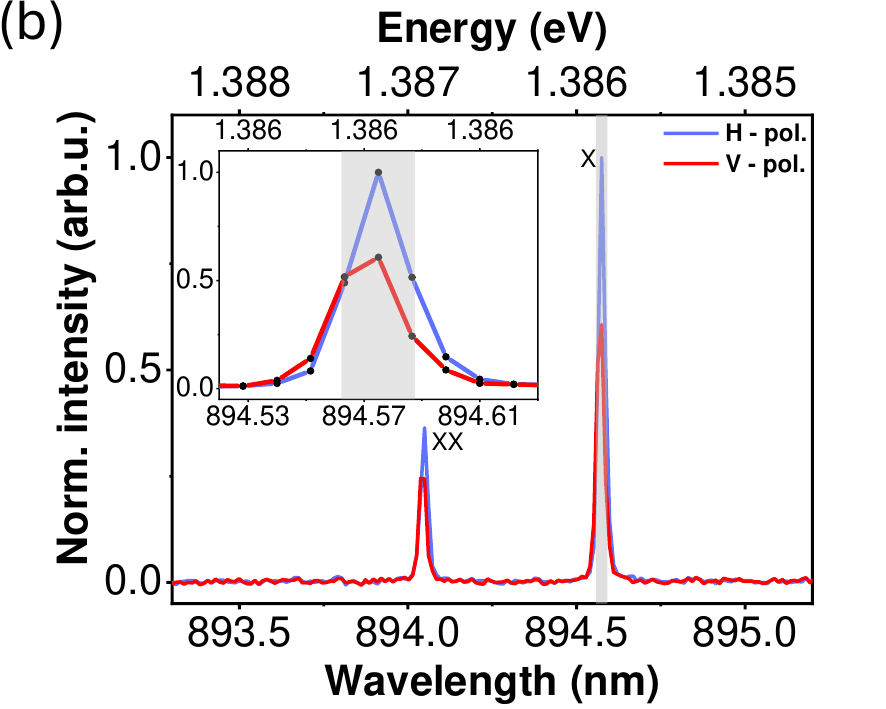}}\\
\subfloat{ \includegraphics[width=0.9\columnwidth]{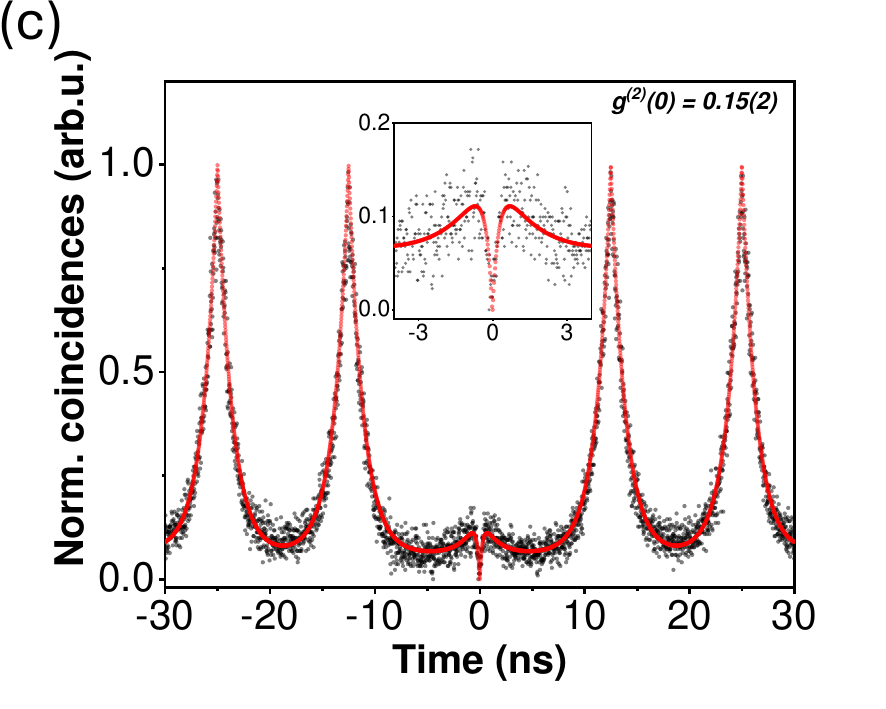}}
\subfloat{ \includegraphics[width=0.9\columnwidth]{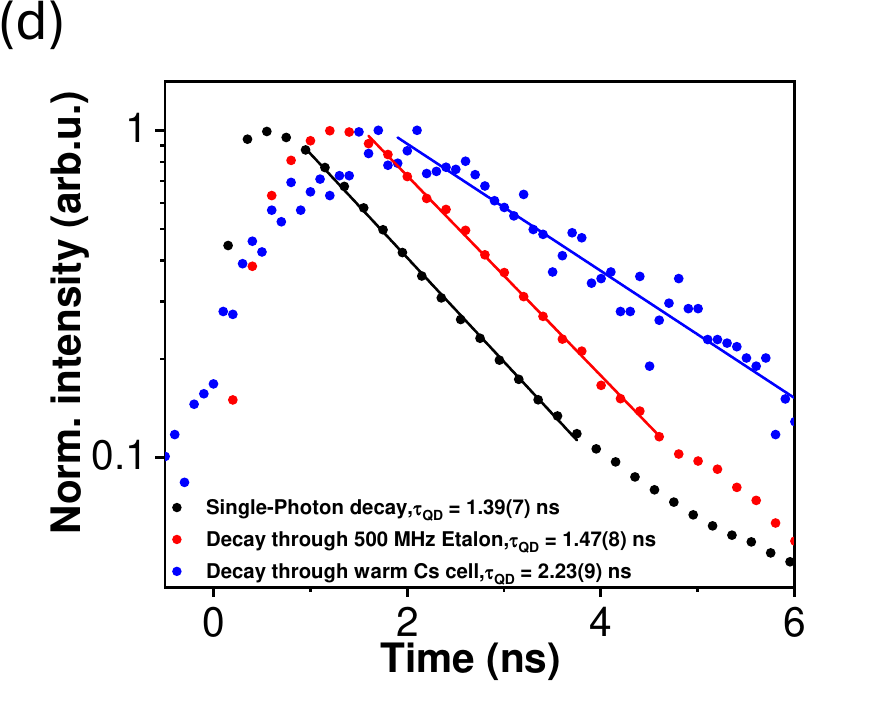}}
    \caption{The QD-hCBG device and its single-photon properties. (a) SEM image (45°-tilted view) of a fabricated QD-hCBG device composed of GaAs membrane after \textit{in-situ} fabrication. (b) Polarization-resolved H-V spectra of the QD integrated into a hCBG device emitting at 17\,K. The inset displays a magnified view of the exciton emission spectra centered around 894.57\,nm aligned with the memory frequency. The gray area represents the spectrometer FWHM resolution, which is about 7.5\,GHz (31\,µeV). (c) Measured and fitted photon autocorrelation function $g^{(2)}(t)$ of the QD at 17\,K. The area-integrated value is $15(2)\%$. The inset shows a magnified view at zero delay. (d) Measured and fitted spectrally filtered emission decay times. The black curve with $\tau_\mathrm{QD}=1.39(7)$\,ns and the red curve with $\tau_\mathrm{QD}=1.47(8)$\,ns correspond to data collected without and with a 500\,MHz etalon filter, respectively. The blue curve with $\tau_\mathrm{QD}=2.23(9)$\,ns represents data collected through both an etalon filter and a $60^\circ$C Cs cell.}
    \label{fig:spectrum}
\end{figure*}

The QD-hCBG device is placed in a continuous flow He-cryostat and excited above the bandgap using a mode-locked Titanium-Sapphire (Ti:Sa) laser emitting ps-pulses with 80\,MHz repetition rate at the central wavelength of 866\,nm. Under non-resonant excitation, the neutral exciton emission of the QD at 4\,K has a wavelength of 894.53\,nm. By heating the sample to 17\,K, the emission is red-detuned by 15(8)\,GHz to spectrally overlap with the Cs-D1 line at 894.57\,nm. To ensure the frequency matching of the QD to the D1 line at a precision beyond the resolution of the spectrometer, a temperature-stabilized Fabry-P{\'e}rot etalon with 500\,MHz linewidth and 25\,GHz free spectral range is used as a fine-grained relative frequency reference. The spectrometer and the etalon are calibrated with a distributed feedback (DFB) laser set on the $F = 4$ to $F^\prime=3$ transition of the Cs D1 line by observing Doppler-free spectroscopy in a Cs cell for absolute frequency reference.

 The QD emission spectrum shown in Fig.~\ref{fig:spectrum}(b) consists of the neutral excitonic (X) and biexcitonic (XX) emission lines with normalized intensity measured using the spectrometer. The blue curve shows the H-polarized component of the fine-structure-split QD emission which is directed to the memory setup, while the red curve shows the V-polarized component which is discarded by polarization filtering.

	Time-resolved second-order correlation measurements were performed to determine the \( g^{(2)}(t) \)-function using a fiber-based Hanbury Brown and Twiss (HBT) configuration \cite{HBT}, see Fig.~\ref{fig:spectrum}(c). The measurements were conducted using superconducting nanowire single-photon detectors (SNSPD), characterized by an instrument response function (IRF) of 93\,ps of the two combined channels in HBT configuration. The analysis of the emission peak corresponding to the Cs transition at the sample temperature of 17\,K yields a \( g^{(2)}(0) \) value of $15(2)\%$ when considering the area under the curve. The inset of the figure provides a magnified view of the \( g^{(2)}(0) \) region. We observe a carrier-recapturing effect near the zero delay which can be attributed to the non-resonant above-band excitation scheme leading to uncontrolled emission time jitter from the non-radiative high-level to s-shell relaxation \cite{Jianweipan2016,Santori_2004}. Though this effect degrades the single-photon purity, we still measure a \( g^{(2)}(0) \) value well below 0.5 which confirms the non-classical nature of the photon emission.	

    \begin{figure*}
    \includegraphics[width=1.8\columnwidth]{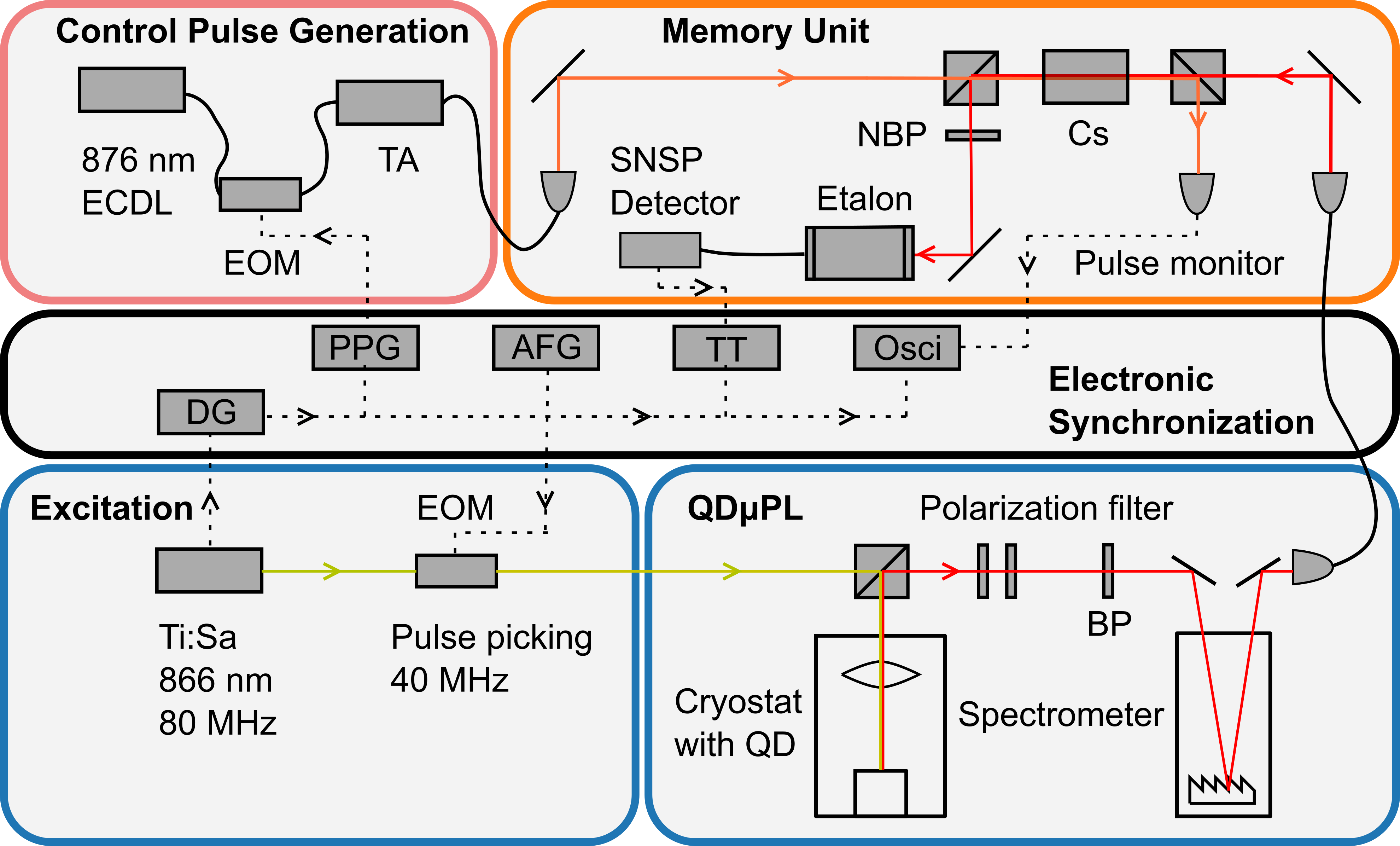}
		\caption{Schematic of the combined experimental source--memory setup. The emitted QD photons are collected with an apochromatic cryogenic objective lens with a high numerical aperture (NA = 0.81). A bandpass filter (BP) (10\,nm FWHM) suppresses the excitation laser. A high-resolution spectrometer (7.5\,GHz, 31\,µeV) is used to monitor the QD emission. A digital delay generator (DG) relays the trigger of the Ti:Sa to a programmable pulse generator (PPG), an arbitrary function generator (AFG), a time tagger (TT) and an oscilloscope (Osci). After retrieval from the memory the QD single photons are frequency filtered with a narrow bandpass (NBP) filter (1\,nm FWHM) and an etalon (500\,MHz FWHM) and detected with a superconducting nanowire single photon detector (SNSPD). TA, tapered amplifier; QD$\mu$PL, QD micro photoluminescence; Cs, cesium vapor cell; EOM, electro-optical modulator;ECDL, external-cavity diode laser.}
  \label{fig:setup}
	\end{figure*}
 
	Figure~\ref{fig:spectrum}(d) shows the results of time-resolved photoluminescence measurements for the QD excitonic emission line with a decay time of $\tau_\mathrm{QD} = 1.39(7)$\,ns. Furthermore, a decay time of $\tau_\mathrm{QD}=2.23(9)$\,ns, was measured when the QD emission traversed both the 60$^\circ$C Cs cell and the 500\,MHz etalon. This longer decay time is consistent with the delay induced by the dispersive interaction of the single photon with the hot Cs vapor inside the cell, as observed in Refs.~\cite{Wildmann.2015, Trotta.2016, Kroh.2019,Bremer:2020,Vural.2021}. Further characterization of optical properties and details on the fabrication of the QD-hCBG device can be found in the Appendix.

\subsection{QD-Memory interface}

 Figure~\ref{fig:setup} shows a schematic of the combined QD-memory setup. The optical configuration described in \ref{sec:QD_characterization} is used to collect and analyze the QD photoluminescence. Following spectral filtering of the excitonic emission through the spectrometer, H-polarized photons are coupled into a single-mode optical fiber and sent to the memory. We do not further filter or manipulate the photons from the QD neither temporally nor spectrally prior to storage in the memory. To demonstrate longer storage times, the excitation rate of the QD-hCBG device is reduced to 40\,MHz with an electro-optical modulator (EOM) acting as a pulse picker. This allows storage and retrieval experiments within a 25\,ns time window after each emission event.

\begin{figure*}
\includegraphics[width=1.8\columnwidth]{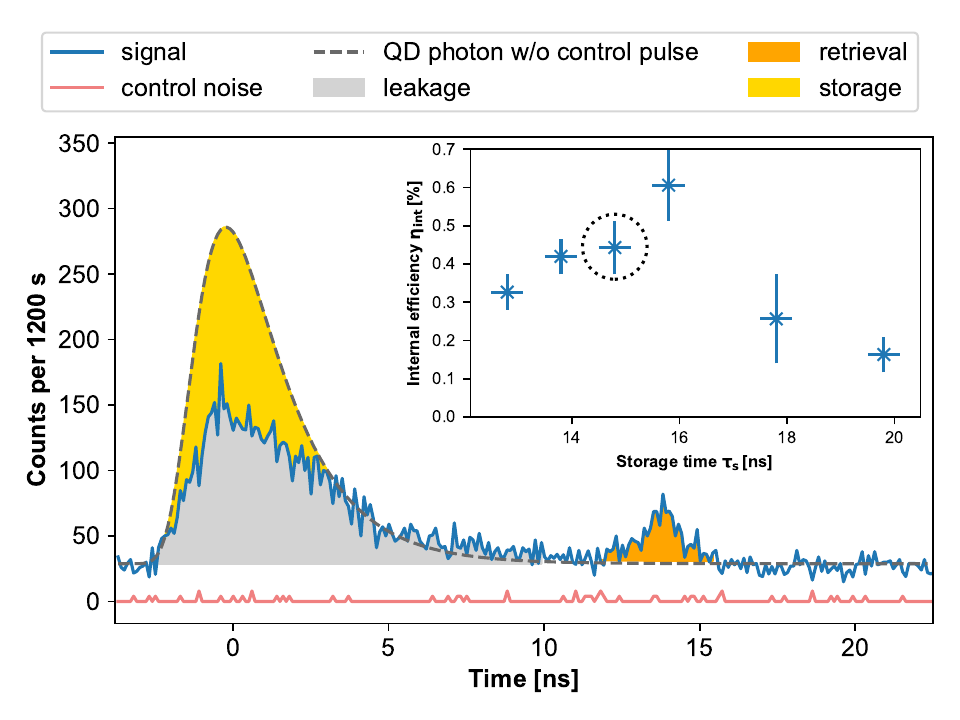}
		\caption{Arrival time histogram of the QD single-photon storage experiment.
        The bin width of the data is 100\,ps. QD photons get stored in the memory (yellow area --storage) and retrieved after the storage time $\tau_\mathrm{s}=13.8(3)$\,ns (orange area --retrieval). The gray dashed line shows the shape of transmitted QD photons without the influence of the control laser. The gray shaded area marks the photons that leak through the memory without being stored. The control laser noise is practically at dark count level of the detector. The constant background of 30 counts per time bin stems from un-timed emission due to the non-resonant excitation of the QD. The inset shows the end-to-end efficiency of the QD single-photon memory for different storage times up to 19.8(3)\,ns. The non-monotonic behavior is expected from the complex spinwave dynamics; due to the beating of multiple addressed hyperfine levels of the storage states as discussed in Refs.~\cite{RanFinkelstein.2018,Maaß.2024}.}
		\label{fig:storage_trace}
	\end{figure*}

The used memory is a warm atomic vapor memory that operates on the Cs D1 line at 895\,nm. We employ a ladder-type configuration \cite{RanFinkelstein.2018,Kaczmarek.2018} that allows for low-noise on-demand storage and retrieval of high-bandwidth single photons. A high-intensity control pulse overlaps with the QD photons within a Cs vapor cell that is heated to 60 $^\circ$C. This results in a spin-orbital coherence of the Cs atoms, the spinwave, which can be mapped back to the initial photon field by re-applying the control pulse. A recent experimental benchmark of the system using attenuated laser pulses has shown that the memory achieves a 1/e storage time of $\approx 32$ ns and up to 15(1)\% internal efficiency at 0.06(2) photons per pulse, while keeping a high signal-to-noise ratio of 830(80) \cite{Maaß.2024}.

The memory exhibits its highest spectral acceptance window of $\mathcal{W}=560(60)$\,MHz at a frequency detuning of $\Delta=-500$\,MHz from the $F = 4$ to $F^\prime=3$ transition of the Cs D1 line. This is larger than the homogeneous linewidth ($\approx 400\,$MHz) of the QD photons. Further characteristics of the memory and a detailed setup sketch can be found in Ref.\cite{Maaß.2024}. To further spectrally suppress uncorrelated background emission from the QD and to enhance the contrast of the retrieved signal, the transmitted light is frequency filtered behind the memory with a temperature stabilized Fabry-Pérot etalon with 500\,MHz linewidth and 25\,GHz free spectral range. The retrieved photons are subsequently detected with an SNSPD.

	\section{Results}
	 The arrival time histogram in Fig.~\ref{fig:storage_trace} shows the storage and retrieval of the QD photons from the memory. The first peak corresponds to the leakage i.e. the photons that have not successfully been read-into the memory due to imperfect temporal mode match, insufficient optical depth and limited availability of control laser power. The second peak corresponds to the number of photons that are retrieved from the memory $N_\text{ret}$ after the storage time $\tau_\text{s}$. To evaluate the efficiency of the memory, we take a reference measurement of the unmodified QD photons that are sent into the memory unit and extract the number of input photons, $N_{\text{input}}$. The end-to-end efficiency of the memory, which includes the setup transmission and filtering losses, is given by: $\eta_\text{e2e}=N_\text{ret}/N_{\text{input}} = 0.026(4)\%$. The number of retrieved photons for the calculation of $\eta_\text{e2e}$ is determined from a Gaussian fit of the retrieval peak and an exponential fit to the QD leakage. The internal efficiency of the storage and retrieval process can be calculated from the setup transmission $T$, which is measured with the calibration laser that is 6\,GHz red-detuned from the $F=4$ to $F^\prime$=3 transition, and the end-to-end efficiency. The memory achieves its maximum internal efficiency at a storage time of 15.8(3)\,ns: $\eta_\text{int}=\eta_\text{e2e} / T=0.6(1)\%$. A detailed loss budget for the setup and the evaluation method can be found in appendix \ref{sec:loss_budget}.
 
 The characterization of the memory with attenuated laser pulses has shown an internal efficiency of 15(4)\%. The main effect that contributes to the lowered efficiency for the storage of QD photons is their large inhomogeneous linewidth of $\delta\nu=5.1(7)$\,GHz compared to the spectral acceptance window of the memory of $\mathcal{W}=560(60)$\,MHz. Only approximately $\mathcal{W}/\delta\nu\approx0.1$ of the QD emission falls within the spectral acceptance window and can be read into the memory. Spectral filtering of the incident QD photons, i.e. putting an etalon in front of the memory, could formally increase the internal efficiency of the memory by a factor 10, without improving the storage and retrieval rate. Taking this into account, the memory efficiency in this work is comparable to Ref.~\cite{Thomas.2024}. Another factor contributing to the reduced efficiency is the mismatch of the temporal mode profiles of the QD photons and the control laser \cite{Gao.2019}. This is estimated to reduce the efficiency by a factor of 0.66. Considering these limiting factors, the memory efficiency in this experiment is consistent with the characterization of the memory in Ref.\cite{Maaß.2024}. The end-to-end efficiency of the setup is stable over a QD temperature range of $\pm$1\,K facilitating long-term operation in future experiments.
		
	The inset of Fig.~\ref{fig:storage_trace} shows the internal efficiency of the memory for different retrieval times. From the Doppler-dephasing of the spinwave one would expect an exponential decay of the efficiency over time. However, we observe a non-monotonic behavior with a maximum at 15.8\,ns storage time. This effect can be attributed to the coupling of the control field to the full hyperfine manifold of the 6D$_{3/2}$ level which results in a beating and rephasing of multiple spinwave contributions. By preparing the atomic ensemble in a stretched state by means of optical Zeeman pumping this effect can be mitigated~\cite{RanFinkelstein.2018}. The accessible retrieval times in this experiment are further limited by the repetition rate of the QD-hCBG SPS and the exponential background during the retrieval window. Technical limitations aside, the memory storage time is fundamentally limited only by the 1/e dephasing time of the spinwave ($\approx$32\,ns) and the lifetime of the 6D$_{3/2}$ storage level.

 \section{Discussion}

 For practical usage of the QD–memory interface a high end-to-end efficiency is crucial. The lack thereof hinders e.g. correlation measurements on the stored and retrieved photons.
 The efficiency of the current setup is predominantly limited by technicalities, namely optical losses and the large inhomogeneous QD linewidth. Significant improvements of the emission properties of the QD-hCBG device, the linewidth and the single photon purity in particular, are expected under resonant or two-photon excitation~\cite{Richard2020} and at a lowered operation temperature. The necessary fine-tuning of the emission wavelength can be achieved with piezo-controlled strain-tuning~\cite{Magdelena2020} or by static electric fields~\cite{wijitpatima2024}. Furthermore, Purcell-engineering of emission in suitably designed hCBGs could further improve the matching of the homogeneous emission linewidth to the memory system. 
 Possible technical measures for improving the efficiency of the memory setup include higher optical depth through incorporation of optical pumping and bandwidth increase through higher control laser power. The storage time of the memory is mostly limited by residual Doppler broadening between the signal and control fields. By means of velocity-selective pumping \cite{Main.2021} or off-resonant dressing of the storage state~\cite{Finkelstein.2021} this limitation can be mitigated. These changes could improve the memory's time-bandwidth product and efficiency, making it practically useful for photon buffering, enhancing the rate of probabilistically generated single photons and allowing for interference experiments between heterogeneous single photon sources.
   
 Already in its current state, this experiment enables on-demand control over the timing of single photons which is a fundamental prerequisite for interconnects between different photonic channels in quantum networks. In such networks the photonic resources will most likely be of heterogeneous nature. Thus, a memory platform that is compatible with a variety of SPS is of utmost importance. In light of our results and other recent publications that demonstrate the compatibility of alkali vapor memories with SPDC sources \cite{Mottola.2020, Buser.2022} and vapor-based sources \cite{Davidson.2023}, room-temperature vapor memories advance as highly promising candidates for routing, buffering and conditioning of quantum information in heterogeneous networks.
  
\section{Conclusion}
In summary, we demonstrate on-demand storage and retrieval of single photons from a QD-hCBG SPS in a room-temperature atomic vapor memory. At variable retrieval times ranging up to 19.8(3)\,ns the memory achieves an internal efficiency of $\eta_\mathrm{int}=0.6(1)\%$ and a time bandwidth product of $B=14(1)$. By providing an interface for real time controllability of the arrival time of single photons from a quantum dot device this work contributes to the development of heterogeneous platforms for buffering and synchronization of quantum network nodes. 
		
	\begin{acknowledgments}
    The authors acknowledge financial support from:\\
		DFG project RE2974/28-1 (448532670); Einstein Foundation (Einstein Research Unit “Perspectives of a quantum digital transformation: Near-term
quantum computational devices and quantum processors”; German Ministry of Science and Education (BMBF) project Q-ToRX (16KISQ040K).\\

		We would like to thank Chirag Palekar, Ching-Wen Shih, Martin von Helversen, Lucas Rickert, Helen Chrzanowski and Leon Messner for fruitful discussions. The authors at KIST (SIP and JDS) acknowledge the partial support from the institutional program of KIST.
	\end{acknowledgments}
	
\appendix

\section{Device design and fabrication}
The nanophotonic device design is crucial for optimizing the interaction between the component geometry and refractive indices, especially when QDs interface with atomic vapors with varying optical properties like different transition linewidths. This optimization process places exact demands on solid-state single-photon emitters' quality and emission characteristics. 

To achieve this optimization, numerical simulations were conducted using the commercial finite element method (FEM) software, JCMsuite \cite{Burger_JCM,JCMWebsite}. By modelling different CBG designs we could evaluate their performance metrics, such as photon extraction efficiency and fabricate the structure accordingly.
The design process was meticulous in accounting for the refractive index dispersion of all relevant materials, including low-temperature GaAs, SiO$_2$, Gold, and Titanium which was essential for ensuring the accurate modelling of the optical properties. The values of the refractive indices have been determined at cryogenic temperature by partial interpolation and extrapolation from the well-established literature to ensure the reliability and accuracy of the simulation \cite{JenkinsOpticalconstants,JphnsonOpticalconsNobelMetal,marpleRIGaAs,RodriguezSiO2constants}.
The device design is based on a hybrid circular Bragg grating (hCBG) resonator structure, which was selected due to its ability to enhance the efficiency of photon extraction from quantum dots \cite{YaoBCBG,Jinliu2019CBG,Nawrath_Michler,Shih:23}. The hCBG design comprises a thin membrane (461\,nm) of GaAs with an InGaAs QD, a central mesa with a diameter of 864\,nm, surrounded by three concentric rings with a period of 540\,nm. To improve the photon collection efficiency and their far-field emission profile, a thin layer of silicon dioxide (250\,nm) was positioned beneath the CBG, followed by 200\,nm of Gold. This hybrid configuration was specifically chosen to redirect downward-propagating light towards the top, thereby improving collection efficiency and yielding a near-Gaussian far-field emission profile \cite{Shih:23}.

The sample fabrication of the single-photon source starts with the growth of the layer structure and quantum dot (QD) formation by the molecular beam epitaxy (MBE) process \cite{MBEgrowth} with two AlGaAs sacrificial layers for the controlled flip-chip process \cite{Wang2019,Jinliu2019CBG}. This technique allows for precise control over the size and composition of the QDs, thereby ensuring compatibility with the subsequent fabrication steps. The subsequent process is followed by the flip-chip gold thermocompression bonding to implement the gold mirror on the backside.

The fabrication process of the QD-hCBG, is based on the in-situ electron beam lithography (i-EBL) technique  \cite{Rodt_2021}. The process commences with cathodoluminescence (CL) spectroscopy at 20 K, which enables the precise mapping of the QD emission characteristics. The initial CL mapping also provides insight into the density of the QDs, which facilitates the identification of low-density areas and finally of individual QDs. 

In the i-EBL process flow, a 300\,nm thick layer of CSAR (AR-P-6200-13) resist is spin-coated onto the flip-chipped QD sample. To achieve precise patterning, a proximity-corrected nano mask design was employed to optimize the size and shape of the final device. Initially, a 20×20\,$\mu\mathrm{m}^2$ area was scanned to remap the sample and accurately locate suitable QDs. The selection of these QDs was performed manually, focusing on those within the target wavelength range of 895(1)\,nm. Subsequently, the identified QDs were subjected to precise electron beam lithography (EBL), whereby CBGs with numerically optimised designs were exposed into the CSAR resist. The CBGs were aligned with great precision to the corresponding QD positions, where the resist is locally inverted, rendering it insoluble during the development step that followed. This precise alignment process, conducted {in situ}, ensured that the CBGs were accurately aligned with the pre-selected QDs, achieving an alignment accuracy of approximately 30–40\,nm \cite{Gschrey.2015,Madigawa.2024}. The precision of the alignment is essentially constrained by the random mechanical drift of the cryostat, which exhibits a drift rate of between $5-9\,$nm/min. 
The non-inverted resist was removed during the development process, leaving the desired CBG structures as an etching mask on the sample surface. The subsequent fabrication step involved the transfer of the CBG profile into the semiconductor material via a highly anisotropic, inductively coupled plasma-reactive ion etching (ICP-RIE) process \cite{ICPRIE}. The etching process resulted in the removal of approximately 461\,nm of GaAs material, including the uncovered quantum dot layer in the designated areas. This led to the formation of GaAs CBGs, each precisely integrated with a single, pre-selected QD. The final structure, as depicted in Fig. 2(a) in the main manuscript, revealed a well-defined device that was optimised as a bright source of single photons.

\section{QD Single-photon characteristics}
\label{sec:QD_characterization}
\begin{figure}
\subfloat{\includegraphics[width=0.9\columnwidth]{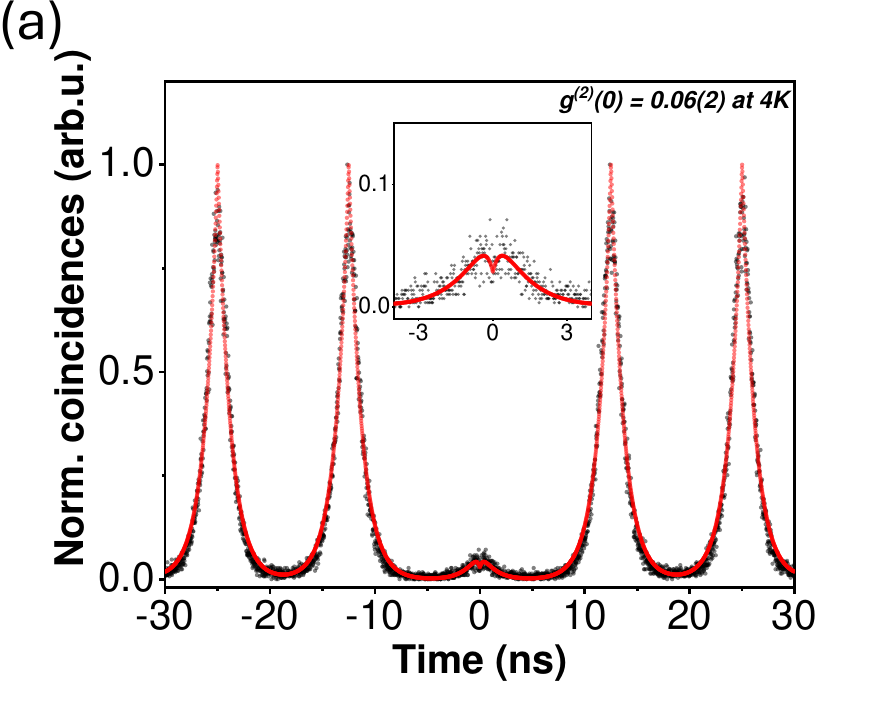}}\\
 \subfloat{\includegraphics[width=0.9\columnwidth]{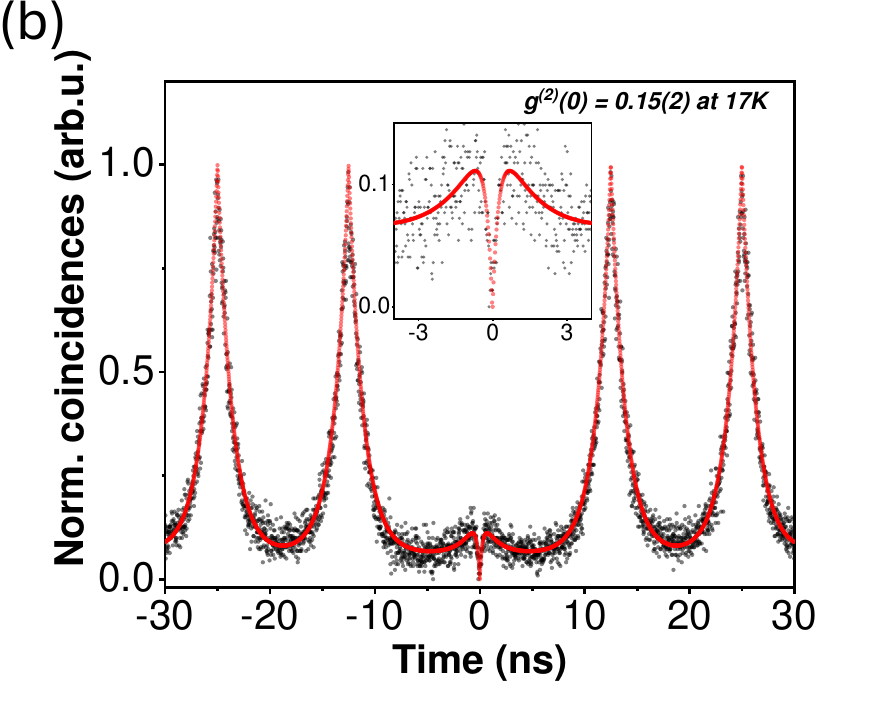}}   
    \caption{Second-order photon autocorrelation function $g^{(2)}(t)$ measured at two sample temperatures. (a) \( g^{(2)}(0) \) of $6(2)\%$ at 4K and (b) \( g^{(2)}(0) \) of $15(2)\%$ at 17K. The Inset image shows the magnified areas at zero delay. }
    \label{g2}
    \end{figure}

The fabricated QD-hCBG devices are characterized to ensure that their performance meets the design specifications. The key parameters evaluated include the photon extraction efficiency (see appendix \ref{sec:loss_budget}), the emission wavelength alignment with the Cs-D1 line with the temperature tuning properties (see section 3), and single-photon emission properties.
Fig.~\ref{g2} shows the measured second-order photon autocorrelation function for two distinct temperatures. The measurement yielded a value \( g^{(2)}(0) \) of $6(2)\%$ at 4\,K and $15(2)\%$ at 17\,K. The fitting area considered under the curve was 12.5\,ns The inset of the figure provides a magnified view of the zero delay. The measured \( g^{(2)}(0) \) values confirm the single-photon nature of the emission even at slightly elevated temperatures.

\section{Frequency tuning with temperature}
\label{sec: Temperature tuning}
The interfacing of the QD source with an atomic quantum memory requires precise spectral alignment with the atomic transition, which demands rigorous control over the emission frequency of the single-photon source. One effective method for attaining the necessary degree of precision is alignment through temperature tuning.
In the present study, we demonstrate that the emission frequency of the QD source can be adjusted over a wide range by varying the temperature. In particular, an adjustment of up to 45\,GHz (0.12\,nm or 186\,$\mu$eV) is observed when the temperature is increased from 4\,K to 21\,K. This illustrates the effectiveness of temperature as a tuning parameter for regulating the QD source's emission characteristics at the cost of slightly deteriorating single-photon purity. (see Appendix B).
 To achieve precise alignment with the atomic memory frequency, the QD emission frequency was fine-tuned at sub-GHz level by raising the device temperature from 4\,K to a maximum of 17\,K which yielded a wavelength tuning of 15\,GHz (0.04\,nm or 62\,$\mu$eV). 
 
\begin{figure}
\subfloat{\includegraphics[width=0.9\columnwidth]{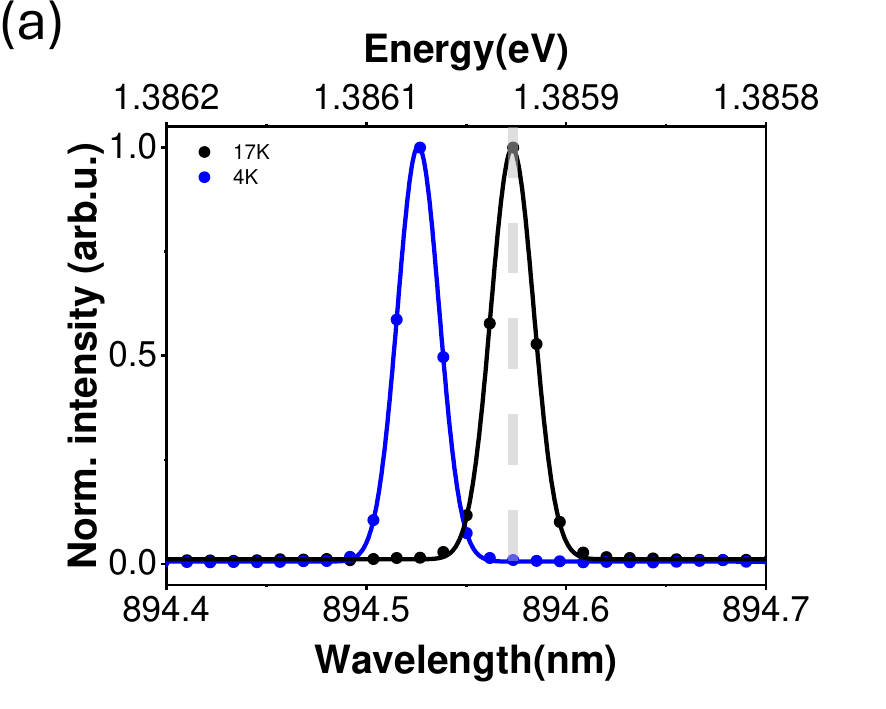}}\\
\subfloat{\includegraphics[width=0.9\columnwidth]{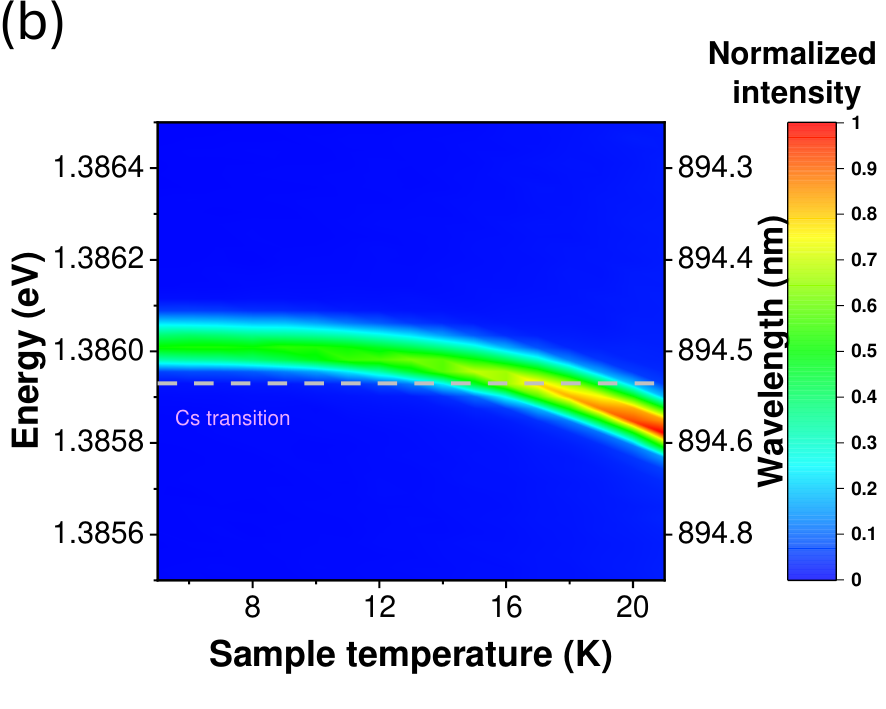}}
   \caption{Normalized $\mu$PL spectra of X emission from the QD-hCBG SPS at 4\,K and 17\,K are presented in (a). In (b) a contour plot of the wavelength tunability of the emission from 4\,K to 21\,K is provided. It can be seen that tunability of up to 45\,GHz is achieved. The gray dashed line corresponds to the Cs-D1 line transition, which matches the QD emission at the sample temperature of 17\,K.}
   \end{figure}
   
\section{Linewidth measurement}
\label{sec: FPI}
The full-width half maximum (FWHM) of the emission is quantified by resolving the emission line through a scanning Fabry-Pérot interferometer (FPI) with a spectral resolution of $150\,$MHz, and a free spectral range (FSR) of $12.3\,$GHz.  Following the spectrometer (1200\,lines/mm grating with 750\,mm focal length) filters the QD emission and the light is directed to the FPI, which is scanned at a rate of $5\,$mHz and $500\,$mVp-p. The distance between the resolved peaks can then be identified with the independently measured free spectral range of the FPI which calibrates the frequency axis. The FWHM of the QD is determined to be $5.1(7)$\,GHz at the sample temperature of 17\,K.

\begin{figure}
    \includegraphics[width=\columnwidth]{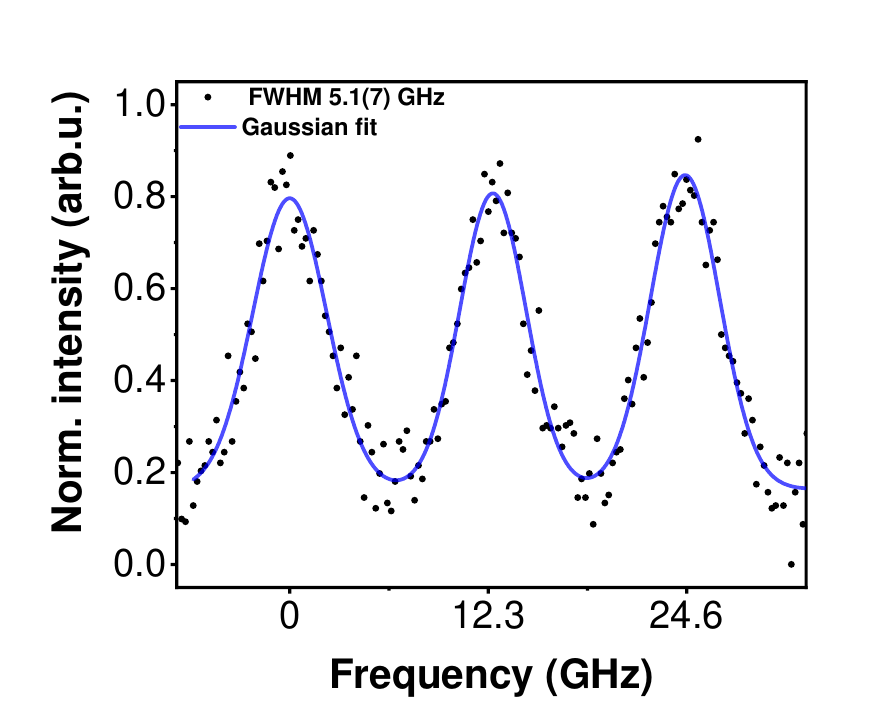}
    \caption{Linewidth measurement with scanning Fabry-Pérot interferometer at 17\,K sample temperature.}
    \label{fig:linewidth}
\end{figure}

\section{Power and polarization dependence}
\label{sec: x and XX}
\begin{figure}
\subfloat{\includegraphics[width=0.9\columnwidth]{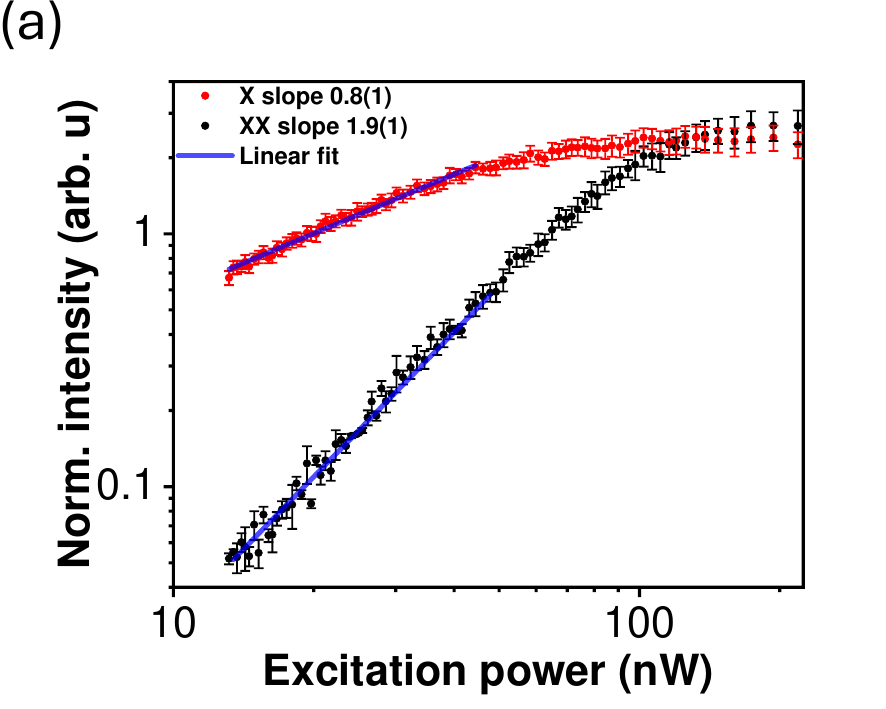}}\\
\subfloat{\includegraphics[width=0.9\columnwidth]{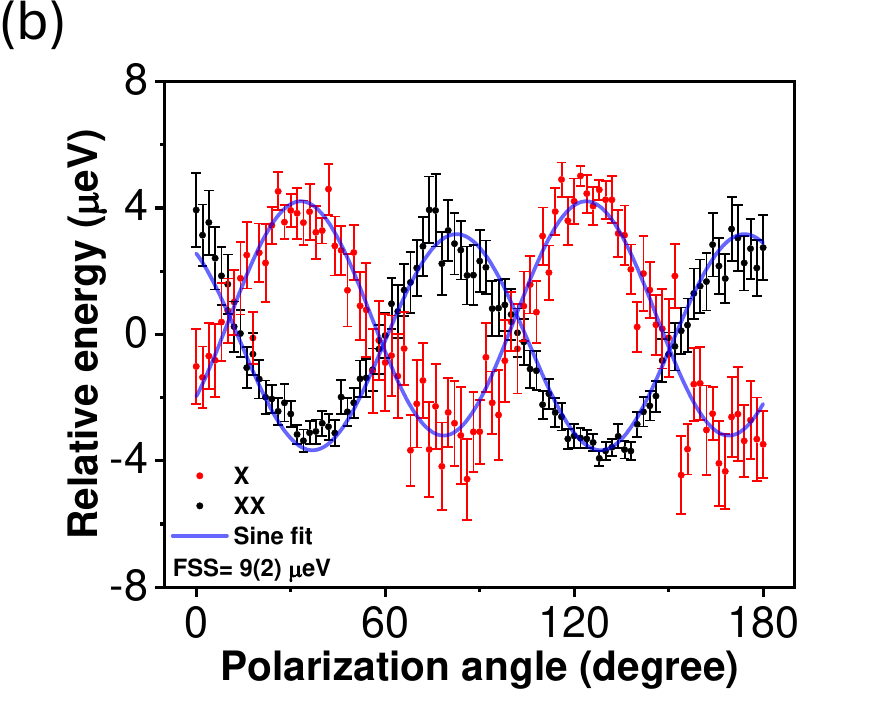}}
    \caption{(a) Power dependent $\mu$PL plot of the integrated intensity versus excitation power in double logarithmic scale, showing the intensity slope variation of the X and XX emission from the QD-hCBG SPS. (b) Polarization-resolved measurement revealing the presence of a fine structure splitting of FSS=$9(2)$ $\mu$eV.}
    \label{fig:powerpol}
\end{figure}

The excitonic complexes of the QD emission lines were determined through power- and polarization- resolved $\mu$PL measurements.
To obtain the power-dependent measurements, the $\mu$PL spectra are recorded at a series of various optical excitation powers. Subsequently, the intensity of the emission lines is selectively fitted, and the resulting values are plotted against the excitation power on a double-logarithmic scale. This allows to extract the power exponent\,(m) from the slope of the linear portion of the plot. Fig.~\ref{fig:powerpol} (a) illustrates the power-dependent $\mu$PL measurements, wherein the fitted emission line demonstrates the extracted power exponent slope $m=0.8(1)$  and $m=1.9(1)$  for the red and data points. respectively. The extracted value of the slope\,(m) allows us to assign emission lines with the linear $(m \sim 1)$ and superlinear $(m>1)$ power dependencies to the exciton\,(x) and biexciton\,(XX), respectively \cite{Finley2001}.  
This assignment is confirmed by additional  polarization dependent µPL studies using a motor-controlled turnable $\lambda/2$-plate and a linear polarizer within the collection path of the $\mu$PL setup. By modifying the angle of the $\lambda/2$-plate we record various polarization-dependent intensities of each emission line. Fig.~\ref{fig:powerpol} (b) illustrates the two polarization components of each emission line as observed along the horizontal (H) and vertical (V) polarization orientation. The spin-related fine structure splitting (FSS) is observed by determining the peak energy differences between two polarization angles \cite{EdigerWarburton2007,Bimberg2009}. The observed energy shift reveals a FSS of $9(2)$ $\mu$eV.

\begin{table*}
\centering
\begin{tabular}{ccc}
\hline
part & transmission/efficiency & count rate [1/s] \\
\hline
Ti:Sa excitation&&$40\cdot10^6$ \\
QD source&(0.40)&$16\cdot 10^6$\\
$\mu$PL optics and spectrometer&0.1&$1.6\cdot 10^6$\\
polarization filtering&(0.5)&$0.8\cdot 10^6$\\
fiber coupling&0.025&$20\cdot 10^3$\\
\hline
memory unit&0.13&\\
etalon&0.5&\\
spectral filtering& 0.66&\\
\hline
\end{tabular}
  \caption{\bf Optical losses in the experimental setup. The line marks the optical losses on the QD setup and on the memory setup, respectively.}
  \label{lossbudget}
\end{table*}

\section{Synchronization}

 The trigger signal of the Ti:Sa Laser serves as the main clock for the whole experiment. A digital delay generator (Highland Technology) relays the trigger signal to the control electronics. In particular, these are the electronic pulse generator that drives the EOM of the control laser, the oscilloscope for pulse monitoring, an arbitrary waveform generator that controls the pulse picking of the Ti:Sa, and the time tagger (QuTools) for single-photon counting. However, since the external trigger rate of the delay generator is limited to 10\,MHz we only detect 1/8 of the trigger events. This results in an overall reduction of the experiment's repetition rate and very long integration times. This is a technical limitation that can easily be overcome with more suitable control electronics.

\section{Loss budget for the optical setup}
\label{sec:loss_budget}
Table \ref{lossbudget} shows the estimated loss budget on the optical setup. The values in parenthesis are assumed or calculated from the measured values.

\section{Efficiency calculations}
To infer the internal efficiency of the storage process from the measured end-to-end efficiency we have to consider the transmission of the retrieved QD photons through the memory. Through the storage process the memory acts as a spectral filter and thereby reduces the linewidth of the retrieved photons to 500\, MHz centered around the spectral acceptance window. We assume that the peak etalon transmission is aligned with the peak of the spectral acceptance window of the memory and therefore with the central frequency of the retrieved photons. Accounting for the mismatch of the 500\,MHz Lorentzian profile of the etalon and the Gaussian profile of the retrieved photons we calculate the filter effect of the etalon to be 0.66. The technical transmission through the memory (0.13) and the etalon (0.5) are measured with a narrow-band (2\,MHz) DFB laser. The overall transmission of the (retrieved) QD photons through the setup is then $T=0.13\cdot0.5\cdot0.66=0.042$. This value is the basis for the calculation of the internal efficiency according to $\eta_{\text{int}} = \eta_{\text{e2e}} / T$.

\bibliography{references}% Produces the bibliography via BibTeX.

\begin{thebibliography}{61}%
\makeatletter
\providecommand \@ifxundefined [1]{%
 \@ifx{#1\undefined}
}%
\providecommand \@ifnum [1]{%
 \ifnum #1\expandafter \@firstoftwo
 \else \expandafter \@secondoftwo
 \fi
}%
\providecommand \@ifx [1]{%
 \ifx #1\expandafter \@firstoftwo
 \else \expandafter \@secondoftwo
 \fi
}%
\providecommand \natexlab [1]{#1}%
\providecommand \enquote  [1]{``#1''}%
\providecommand \bibnamefont  [1]{#1}%
\providecommand \bibfnamefont [1]{#1}%
\providecommand \citenamefont [1]{#1}%
\providecommand \href@noop [0]{\@secondoftwo}%
\providecommand \href [0]{\begingroup \@sanitize@url \@href}%
\providecommand \@href[1]{\@@startlink{#1}\@@href}%
\providecommand \@@href[1]{\endgroup#1\@@endlink}%
\providecommand \@sanitize@url [0]{\catcode `\\12\catcode `\$12\catcode `\&12\catcode `\#12\catcode `\^12\catcode `\_12\catcode `\%12\relax}%
\providecommand \@@startlink[1]{}%
\providecommand \@@endlink[0]{}%
\providecommand \url  [0]{\begingroup\@sanitize@url \@url }%
\providecommand \@url [1]{\endgroup\@href {#1}{\urlprefix }}%
\providecommand \urlprefix  [0]{URL }%
\providecommand \Eprint [0]{\href }%
\providecommand \doibase [0]{https://doi.org/}%
\providecommand \selectlanguage [0]{\@gobble}%
\providecommand \bibinfo  [0]{\@secondoftwo}%
\providecommand \bibfield  [0]{\@secondoftwo}%
\providecommand \translation [1]{[#1]}%
\providecommand \BibitemOpen [0]{}%
\providecommand \bibitemStop [0]{}%
\providecommand \bibitemNoStop [0]{.\EOS\space}%
\providecommand \EOS [0]{\spacefactor3000\relax}%
\providecommand \BibitemShut  [1]{\csname bibitem#1\endcsname}%
\let\auto@bib@innerbib\@empty
%</preamble>
\bibitem [{\citenamefont {Kimble}(2008)}]{Kimble.2008}%
  \BibitemOpen
  \bibfield  {author} {\bibinfo {author} {\bibfnamefont {H.~J.}\ \bibnamefont {Kimble}},\ }\bibfield  {title} {\bibinfo {title} {The quantum internet},\ }\href {https://doi.org/10.1038/nature07127} {\bibfield  {journal} {\bibinfo  {journal} {Nature}\ }\textbf {\bibinfo {volume} {453}},\ \bibinfo {pages} {1023} (\bibinfo {year} {2008})}\BibitemShut {NoStop}%
\bibitem [{\citenamefont {Duan}\ \emph {et~al.}(2001)\citenamefont {Duan}, \citenamefont {Lukin}, \citenamefont {Cirac},\ and\ \citenamefont {Zoller}}]{Duan.2001}%
  \BibitemOpen
  \bibfield  {author} {\bibinfo {author} {\bibfnamefont {L.-M.}\ \bibnamefont {Duan}}, \bibinfo {author} {\bibfnamefont {M.~D.}\ \bibnamefont {Lukin}}, \bibinfo {author} {\bibfnamefont {J.~I.}\ \bibnamefont {Cirac}},\ and\ \bibinfo {author} {\bibfnamefont {P.}~\bibnamefont {Zoller}},\ }\bibfield  {title} {\bibinfo {title} {Long-distance quantum communication with atomic ensembles and linear optics},\ }\href {https://doi.org/10.1038/35106500} {\bibfield  {journal} {\bibinfo  {journal} {Nature}\ }\textbf {\bibinfo {volume} {414}},\ \bibinfo {pages} {413} (\bibinfo {year} {2001})}\BibitemShut {NoStop}%
\bibitem [{\citenamefont {Wang}\ \emph {et~al.}(2022)\citenamefont {Wang}, \citenamefont {Craddock}, \citenamefont {Sekelsky}, \citenamefont {Flament},\ and\ \citenamefont {Namazi}}]{Mehdi.2022}%
  \BibitemOpen
  \bibfield  {author} {\bibinfo {author} {\bibfnamefont {Y.}~\bibnamefont {Wang}}, \bibinfo {author} {\bibfnamefont {A.~N.}\ \bibnamefont {Craddock}}, \bibinfo {author} {\bibfnamefont {R.}~\bibnamefont {Sekelsky}}, \bibinfo {author} {\bibfnamefont {M.}~\bibnamefont {Flament}},\ and\ \bibinfo {author} {\bibfnamefont {M.}~\bibnamefont {Namazi}},\ }\bibfield  {title} {\bibinfo {title} {Field-deployable quantum memory for quantum networking},\ }\href {https://doi.org/10.1103/PhysRevApplied.18.044058} {\bibfield  {journal} {\bibinfo  {journal} {Phys. Rev. Appl.}\ }\textbf {\bibinfo {volume} {18}},\ \bibinfo {pages} {044058} (\bibinfo {year} {2022})}\BibitemShut {NoStop}%
\bibitem [{\citenamefont {Jutisz}\ \emph {et~al.}(2024)\citenamefont {Jutisz}, \citenamefont {Erl}, \citenamefont {Wolters}, \citenamefont {G{\"u}ndo{\u{g}}an},\ and\ \citenamefont {Krutzik}}]{Jutisz.2024}%
  \BibitemOpen
  \bibfield  {author} {\bibinfo {author} {\bibfnamefont {M.}~\bibnamefont {Jutisz}}, \bibinfo {author} {\bibfnamefont {A.}~\bibnamefont {Erl}}, \bibinfo {author} {\bibfnamefont {J.}~\bibnamefont {Wolters}}, \bibinfo {author} {\bibfnamefont {M.}~\bibnamefont {G{\"u}ndo{\u{g}}an}},\ and\ \bibinfo {author} {\bibfnamefont {M.}~\bibnamefont {Krutzik}},\ }\href {http://arxiv.org/pdf/2410.21209} {\bibinfo {title} {Standalone mobile quantum memory system}} (\bibinfo {year} {2024})\BibitemShut {NoStop}%
\bibitem [{\citenamefont {Heindel}\ \emph {et~al.}(2023)\citenamefont {Heindel}, \citenamefont {Kim}, \citenamefont {Gregersen}, \citenamefont {Rastelli},\ and\ \citenamefont {Reitzenstein}}]{Heindel2023}%
  \BibitemOpen
  \bibfield  {author} {\bibinfo {author} {\bibfnamefont {T.}~\bibnamefont {Heindel}}, \bibinfo {author} {\bibfnamefont {J.-H.}\ \bibnamefont {Kim}}, \bibinfo {author} {\bibfnamefont {N.}~\bibnamefont {Gregersen}}, \bibinfo {author} {\bibfnamefont {A.}~\bibnamefont {Rastelli}},\ and\ \bibinfo {author} {\bibfnamefont {S.}~\bibnamefont {Reitzenstein}},\ }\bibfield  {title} {\bibinfo {title} {Quantum dots for photonic quantum information technology},\ }\href {https://doi.org/10.1364/aop.490091} {\bibfield  {journal} {\bibinfo  {journal} {Advances in Optics and Photonics}\ }\textbf {\bibinfo {volume} {15}},\ \bibinfo {pages} {613} (\bibinfo {year} {2023})}\BibitemShut {NoStop}%
\bibitem [{\citenamefont {Senellart}\ \emph {et~al.}(2017)\citenamefont {Senellart}, \citenamefont {Solomon},\ and\ \citenamefont {White}}]{Pascalesenellart}%
  \BibitemOpen
  \bibfield  {author} {\bibinfo {author} {\bibfnamefont {P.}~\bibnamefont {Senellart}}, \bibinfo {author} {\bibfnamefont {G.}~\bibnamefont {Solomon}},\ and\ \bibinfo {author} {\bibfnamefont {A.}~\bibnamefont {White}},\ }\bibfield  {title} {\bibinfo {title} {High-performance semiconductor quantum-dot single-photon sources},\ }\bibfield  {journal} {\bibinfo  {journal} {Nature Nanotechnology}\ }\textbf {\bibinfo {volume} {12}},\ \href {https://doi.org/10.1038/nnano.2017.218} {10.1038/nnano.2017.218} (\bibinfo {year} {2017})\BibitemShut {NoStop}%
\bibitem [{\citenamefont {Yu}\ \emph {et~al.}(2023)\citenamefont {Yu}, \citenamefont {Liu}, \citenamefont {Lee}, \citenamefont {Michler}, \citenamefont {Reitzenstein}, \citenamefont {Srinivasan}, \citenamefont {Waks},\ and\ \citenamefont {Liu}}]{Yu2023}%
  \BibitemOpen
  \bibfield  {author} {\bibinfo {author} {\bibfnamefont {Y.}~\bibnamefont {Yu}}, \bibinfo {author} {\bibfnamefont {S.}~\bibnamefont {Liu}}, \bibinfo {author} {\bibfnamefont {C.-M.}\ \bibnamefont {Lee}}, \bibinfo {author} {\bibfnamefont {P.}~\bibnamefont {Michler}}, \bibinfo {author} {\bibfnamefont {S.}~\bibnamefont {Reitzenstein}}, \bibinfo {author} {\bibfnamefont {K.}~\bibnamefont {Srinivasan}}, \bibinfo {author} {\bibfnamefont {E.}~\bibnamefont {Waks}},\ and\ \bibinfo {author} {\bibfnamefont {J.}~\bibnamefont {Liu}},\ }\bibfield  {title} {\bibinfo {title} {Telecom-band quantum dot technologies for long-distance quantum networks},\ }\href {https://doi.org/10.1038/s41565-023-01528-7} {\bibfield  {journal} {\bibinfo  {journal} {Nature Nanotechnology}\ }\textbf {\bibinfo {volume} {18}},\ \bibinfo {pages} {1389–1400} (\bibinfo {year} {2023})}\BibitemShut {NoStop}%
\bibitem [{\citenamefont {Vajner}\ \emph {et~al.}(2022)\citenamefont {Vajner}, \citenamefont {Rickert}, \citenamefont {Gao}, \citenamefont {Kaymazlar},\ and\ \citenamefont {Heindel}}]{DanielAQTReview}%
  \BibitemOpen
  \bibfield  {author} {\bibinfo {author} {\bibfnamefont {D.~A.}\ \bibnamefont {Vajner}}, \bibinfo {author} {\bibfnamefont {L.}~\bibnamefont {Rickert}}, \bibinfo {author} {\bibfnamefont {T.}~\bibnamefont {Gao}}, \bibinfo {author} {\bibfnamefont {K.}~\bibnamefont {Kaymazlar}},\ and\ \bibinfo {author} {\bibfnamefont {T.}~\bibnamefont {Heindel}},\ }\bibfield  {title} {\bibinfo {title} {Quantum communication using semiconductor quantum dots},\ }\href {https://doi.org/https://doi.org/10.1002/qute.202100116} {\bibfield  {journal} {\bibinfo  {journal} {Advanced Quantum Technologies}\ }\textbf {\bibinfo {volume} {5}},\ \bibinfo {pages} {2100116} (\bibinfo {year} {2022})},\ \Eprint {https://arxiv.org/abs/https://onlinelibrary.wiley.com/doi/pdf/10.1002/qute.202100116} {https://onlinelibrary.wiley.com/doi/pdf/10.1002/qute.202100116} \BibitemShut {NoStop}%
\bibitem [{\citenamefont {Lo}\ \emph {et~al.}(2014)\citenamefont {Lo}, \citenamefont {Curty},\ and\ \citenamefont {Tamaki}}]{HKLo_nature20214}%
  \BibitemOpen
  \bibfield  {author} {\bibinfo {author} {\bibfnamefont {H.-K.}\ \bibnamefont {Lo}}, \bibinfo {author} {\bibfnamefont {M.}~\bibnamefont {Curty}},\ and\ \bibinfo {author} {\bibfnamefont {K.}~\bibnamefont {Tamaki}},\ }\bibfield  {title} {\bibinfo {title} {Secure quantum key distribution},\ }\bibfield  {journal} {\bibinfo  {journal} {Nature Photonics}\ }\textbf {\bibinfo {volume} {8}},\ \href {https://doi.org/10.1038/nphoton.2014.149} {10.1038/nphoton.2014.149} (\bibinfo {year} {2014})\BibitemShut {NoStop}%
\bibitem [{\citenamefont {Basset}\ \emph {et~al.}(2021)\citenamefont {Basset}, \citenamefont {Valeri}, \citenamefont {Roccia}, \citenamefont {Muredda}, \citenamefont {Poderini}, \citenamefont {Neuwirth}, \citenamefont {Spagnolo}, \citenamefont {Rota}, \citenamefont {Carvacho}, \citenamefont {Sciarrino},\ and\ \citenamefont {Trotta}}]{Trotta2021}%
  \BibitemOpen
  \bibfield  {author} {\bibinfo {author} {\bibfnamefont {F.~B.}\ \bibnamefont {Basset}}, \bibinfo {author} {\bibfnamefont {M.}~\bibnamefont {Valeri}}, \bibinfo {author} {\bibfnamefont {E.}~\bibnamefont {Roccia}}, \bibinfo {author} {\bibfnamefont {V.}~\bibnamefont {Muredda}}, \bibinfo {author} {\bibfnamefont {D.}~\bibnamefont {Poderini}}, \bibinfo {author} {\bibfnamefont {J.}~\bibnamefont {Neuwirth}}, \bibinfo {author} {\bibfnamefont {N.}~\bibnamefont {Spagnolo}}, \bibinfo {author} {\bibfnamefont {M.~B.}\ \bibnamefont {Rota}}, \bibinfo {author} {\bibfnamefont {G.}~\bibnamefont {Carvacho}}, \bibinfo {author} {\bibfnamefont {F.}~\bibnamefont {Sciarrino}},\ and\ \bibinfo {author} {\bibfnamefont {R.}~\bibnamefont {Trotta}},\ }\bibfield  {title} {\bibinfo {title} {Quantum key distribution with entangled photons generated on demand by a quantum dot},\ }\href {https://doi.org/10.1126/sciadv.abe6379} {\bibfield  {journal} {\bibinfo  {journal} {Science Advances}\ }\textbf {\bibinfo {volume} {7}},\ \bibinfo {pages}
  {eabe6379} (\bibinfo {year} {2021})},\ \Eprint {https://arxiv.org/abs/https://www.science.org/doi/pdf/10.1126/sciadv.abe6379} {https://www.science.org/doi/pdf/10.1126/sciadv.abe6379} \BibitemShut {NoStop}%
\bibitem [{\citenamefont {K{\'o}m{\'a}r}\ \emph {et~al.}(2014)\citenamefont {K{\'o}m{\'a}r}, \citenamefont {Kessler}, \citenamefont {Bishof}, \citenamefont {Jiang}, \citenamefont {S{\o}rensen}, \citenamefont {Ye},\ and\ \citenamefont {Lukin}}]{Komar.2014}%
  \BibitemOpen
  \bibfield  {author} {\bibinfo {author} {\bibfnamefont {P.}~\bibnamefont {K{\'o}m{\'a}r}}, \bibinfo {author} {\bibfnamefont {E.~M.}\ \bibnamefont {Kessler}}, \bibinfo {author} {\bibfnamefont {M.}~\bibnamefont {Bishof}}, \bibinfo {author} {\bibfnamefont {L.}~\bibnamefont {Jiang}}, \bibinfo {author} {\bibfnamefont {A.~S.}\ \bibnamefont {S{\o}rensen}}, \bibinfo {author} {\bibfnamefont {J.}~\bibnamefont {Ye}},\ and\ \bibinfo {author} {\bibfnamefont {M.~D.}\ \bibnamefont {Lukin}},\ }\bibfield  {title} {\bibinfo {title} {A quantum network of clocks},\ }\href {https://doi.org/10.1038/nphys3000} {\bibfield  {journal} {\bibinfo  {journal} {Nature Physics}\ }\textbf {\bibinfo {volume} {10}},\ \bibinfo {pages} {582} (\bibinfo {year} {2014})}\BibitemShut {NoStop}%
\bibitem [{\citenamefont {Zhao}\ \emph {et~al.}(2021)\citenamefont {Zhao}, \citenamefont {Zhang}, \citenamefont {Liu}, \citenamefont {Guan}, \citenamefont {Zhang}, \citenamefont {Li}, \citenamefont {Bai}, \citenamefont {Li}, \citenamefont {Liu}, \citenamefont {You}, \citenamefont {Zhang}, \citenamefont {Fan}, \citenamefont {Xu}, \citenamefont {Zhang},\ and\ \citenamefont {Pan}}]{Zhao.2021}%
  \BibitemOpen
  \bibfield  {author} {\bibinfo {author} {\bibfnamefont {S.-R.}\ \bibnamefont {Zhao}}, \bibinfo {author} {\bibfnamefont {Y.-Z.}\ \bibnamefont {Zhang}}, \bibinfo {author} {\bibfnamefont {W.-Z.}\ \bibnamefont {Liu}}, \bibinfo {author} {\bibfnamefont {J.-Y.}\ \bibnamefont {Guan}}, \bibinfo {author} {\bibfnamefont {W.}~\bibnamefont {Zhang}}, \bibinfo {author} {\bibfnamefont {C.-L.}\ \bibnamefont {Li}}, \bibinfo {author} {\bibfnamefont {B.}~\bibnamefont {Bai}}, \bibinfo {author} {\bibfnamefont {M.-H.}\ \bibnamefont {Li}}, \bibinfo {author} {\bibfnamefont {Y.}~\bibnamefont {Liu}}, \bibinfo {author} {\bibfnamefont {L.}~\bibnamefont {You}}, \bibinfo {author} {\bibfnamefont {J.}~\bibnamefont {Zhang}}, \bibinfo {author} {\bibfnamefont {J.}~\bibnamefont {Fan}}, \bibinfo {author} {\bibfnamefont {F.}~\bibnamefont {Xu}}, \bibinfo {author} {\bibfnamefont {Q.}~\bibnamefont {Zhang}},\ and\ \bibinfo {author} {\bibfnamefont {J.-W.}\ \bibnamefont {Pan}},\ }\bibfield  {title} {\bibinfo {title} {Field demonstration of distributed
  quantum sensing without post-selection},\ }\href {https://doi.org/10.1103/PhysRevX.11.031009} {\bibfield  {journal} {\bibinfo  {journal} {Phys. Rev. X}\ }\textbf {\bibinfo {volume} {11}},\ \bibinfo {pages} {031009} (\bibinfo {year} {2021})}\BibitemShut {NoStop}%
\bibitem [{\citenamefont {Nichol}\ \emph {et~al.}(2022)\citenamefont {Nichol}, \citenamefont {Srinivas}, \citenamefont {Nadlinger}, \citenamefont {Drmota}, \citenamefont {Main}, \citenamefont {Araneda}, \citenamefont {Ballance},\ and\ \citenamefont {Lucas}}]{Nichol.2022}%
  \BibitemOpen
  \bibfield  {author} {\bibinfo {author} {\bibfnamefont {B.~C.}\ \bibnamefont {Nichol}}, \bibinfo {author} {\bibfnamefont {R.}~\bibnamefont {Srinivas}}, \bibinfo {author} {\bibfnamefont {D.~P.}\ \bibnamefont {Nadlinger}}, \bibinfo {author} {\bibfnamefont {P.}~\bibnamefont {Drmota}}, \bibinfo {author} {\bibfnamefont {D.}~\bibnamefont {Main}}, \bibinfo {author} {\bibfnamefont {G.}~\bibnamefont {Araneda}}, \bibinfo {author} {\bibfnamefont {C.~J.}\ \bibnamefont {Ballance}},\ and\ \bibinfo {author} {\bibfnamefont {D.~M.}\ \bibnamefont {Lucas}},\ }\bibfield  {title} {\bibinfo {title} {An elementary quantum network of entangled optical atomic clocks},\ }\href {https://doi.org/10.1038/s41586-022-05088-z} {\bibfield  {journal} {\bibinfo  {journal} {Nature}\ }\textbf {\bibinfo {volume} {609}},\ \bibinfo {pages} {689} (\bibinfo {year} {2022})}\BibitemShut {NoStop}%
\bibitem [{\citenamefont {Wehner}\ \emph {et~al.}(2018)\citenamefont {Wehner}, \citenamefont {Elkouss},\ and\ \citenamefont {Hanson}}]{Wehner2018DistributedQC}%
  \BibitemOpen
  \bibfield  {author} {\bibinfo {author} {\bibfnamefont {S.}~\bibnamefont {Wehner}}, \bibinfo {author} {\bibfnamefont {D.}~\bibnamefont {Elkouss}},\ and\ \bibinfo {author} {\bibfnamefont {R.}~\bibnamefont {Hanson}},\ }\bibfield  {title} {\bibinfo {title} {Quantum internet: A vision for the road ahead},\ }\href {https://doi.org/10.1126/science.aam9288} {\bibfield  {journal} {\bibinfo  {journal} {Science}\ }\textbf {\bibinfo {volume} {362}},\ \bibinfo {pages} {eaam9288} (\bibinfo {year} {2018})},\ \Eprint {https://arxiv.org/abs/https://www.science.org/doi/pdf/10.1126/science.aam9288} {https://www.science.org/doi/pdf/10.1126/science.aam9288} \BibitemShut {NoStop}%
\bibitem [{\citenamefont {Zhai}\ \emph {et~al.}(2020{\natexlab{a}})\citenamefont {Zhai}, \citenamefont {Löbl}, \citenamefont {Nguyen}, \citenamefont {Javadi}, \citenamefont {Spinnler},\ and\ \citenamefont {Warburton}}]{Warburton2020}%
  \BibitemOpen
  \bibfield  {author} {\bibinfo {author} {\bibfnamefont {L.}~\bibnamefont {Zhai}}, \bibinfo {author} {\bibfnamefont {M.~C.}\ \bibnamefont {Löbl}}, \bibinfo {author} {\bibfnamefont {G.~N.}\ \bibnamefont {Nguyen}}, \bibinfo {author} {\bibfnamefont {A.}~\bibnamefont {Javadi}}, \bibinfo {author} {\bibfnamefont {C.}~\bibnamefont {Spinnler}},\ and\ \bibinfo {author} {\bibfnamefont {R.~J.}\ \bibnamefont {Warburton}},\ }\bibfield  {title} {\bibinfo {title} {Low-noise {G}a{A}s quantum dots for quantum photonics},\ }\bibfield  {journal} {\bibinfo  {journal} {Nature Communications}\ }\textbf {\bibinfo {volume} {11}},\ \href {https://doi.org/10.1038/s41467-020-18625-z} {10.1038/s41467-020-18625-z} (\bibinfo {year} {2020}{\natexlab{a}})\BibitemShut {NoStop}%
\bibitem [{\citenamefont {Akopian}\ \emph {et~al.}(2011)\citenamefont {Akopian}, \citenamefont {Wang}, \citenamefont {Rastelli}, \citenamefont {Schmidt},\ and\ \citenamefont {Zwiller}}]{Akopian.2011}%
  \BibitemOpen
  \bibfield  {author} {\bibinfo {author} {\bibfnamefont {N.}~\bibnamefont {Akopian}}, \bibinfo {author} {\bibfnamefont {L.}~\bibnamefont {Wang}}, \bibinfo {author} {\bibfnamefont {A.}~\bibnamefont {Rastelli}}, \bibinfo {author} {\bibfnamefont {O.~G.}\ \bibnamefont {Schmidt}},\ and\ \bibinfo {author} {\bibfnamefont {V.}~\bibnamefont {Zwiller}},\ }\bibfield  {title} {\bibinfo {title} {Hybrid semiconductor-atomic interface: slowing down single photons from a quantum dot},\ }\href {https://doi.org/10.1038/nphoton.2011.16} {\bibfield  {journal} {\bibinfo  {journal} {Nature Photonics}\ }\textbf {\bibinfo {volume} {5}},\ \bibinfo {pages} {230} (\bibinfo {year} {2011})}\BibitemShut {NoStop}%
\bibitem [{\citenamefont {Wildmann}\ \emph {et~al.}(2015)\citenamefont {Wildmann}, \citenamefont {Trotta}, \citenamefont {Mart\'{\i}n-S\'anchez}, \citenamefont {Zallo}, \citenamefont {O'Steen}, \citenamefont {Schmidt},\ and\ \citenamefont {Rastelli}}]{Wildmann.2015}%
  \BibitemOpen
  \bibfield  {author} {\bibinfo {author} {\bibfnamefont {J.~S.}\ \bibnamefont {Wildmann}}, \bibinfo {author} {\bibfnamefont {R.}~\bibnamefont {Trotta}}, \bibinfo {author} {\bibfnamefont {J.}~\bibnamefont {Mart\'{\i}n-S\'anchez}}, \bibinfo {author} {\bibfnamefont {E.}~\bibnamefont {Zallo}}, \bibinfo {author} {\bibfnamefont {M.}~\bibnamefont {O'Steen}}, \bibinfo {author} {\bibfnamefont {O.~G.}\ \bibnamefont {Schmidt}},\ and\ \bibinfo {author} {\bibfnamefont {A.}~\bibnamefont {Rastelli}},\ }\bibfield  {title} {\bibinfo {title} {Atomic clouds as spectrally selective and tunable delay lines for single photons from quantum dots},\ }\href {https://doi.org/10.1103/PhysRevB.92.235306} {\bibfield  {journal} {\bibinfo  {journal} {Phys. Rev. B}\ }\textbf {\bibinfo {volume} {92}},\ \bibinfo {pages} {235306} (\bibinfo {year} {2015})}\BibitemShut {NoStop}%
\bibitem [{\citenamefont {Kroh}\ \emph {et~al.}(2019)\citenamefont {Kroh}, \citenamefont {Wolters}, \citenamefont {Ahlrichs}, \citenamefont {Schell}, \citenamefont {Thoma}, \citenamefont {Reitzenstein}, \citenamefont {Wildmann}, \citenamefont {Zallo}, \citenamefont {Trotta}, \citenamefont {Rastelli}, \citenamefont {Schmidt},\ and\ \citenamefont {Benson}}]{Kroh.2019}%
  \BibitemOpen
  \bibfield  {author} {\bibinfo {author} {\bibfnamefont {T.}~\bibnamefont {Kroh}}, \bibinfo {author} {\bibfnamefont {J.}~\bibnamefont {Wolters}}, \bibinfo {author} {\bibfnamefont {A.}~\bibnamefont {Ahlrichs}}, \bibinfo {author} {\bibfnamefont {A.~W.}\ \bibnamefont {Schell}}, \bibinfo {author} {\bibfnamefont {A.}~\bibnamefont {Thoma}}, \bibinfo {author} {\bibfnamefont {S.}~\bibnamefont {Reitzenstein}}, \bibinfo {author} {\bibfnamefont {J.~S.}\ \bibnamefont {Wildmann}}, \bibinfo {author} {\bibfnamefont {E.}~\bibnamefont {Zallo}}, \bibinfo {author} {\bibfnamefont {R.}~\bibnamefont {Trotta}}, \bibinfo {author} {\bibfnamefont {A.}~\bibnamefont {Rastelli}}, \bibinfo {author} {\bibfnamefont {O.~G.}\ \bibnamefont {Schmidt}},\ and\ \bibinfo {author} {\bibfnamefont {O.}~\bibnamefont {Benson}},\ }\bibfield  {title} {\bibinfo {title} {Slow and fast single photons from a quantum dot interacting with the excited state hyperfine structure of the cesium {D}1-line},\ }\href {https://doi.org/10.1038/s41598-019-50062-x}
  {\bibfield  {journal} {\bibinfo  {journal} {Scientific Reports}\ }\textbf {\bibinfo {volume} {9}},\ \bibinfo {pages} {13728} (\bibinfo {year} {2019})}\BibitemShut {NoStop}%
\bibitem [{\citenamefont {Bremer}\ \emph {et~al.}(2020)\citenamefont {Bremer}, \citenamefont {Fischbach}, \citenamefont {Park}, \citenamefont {Rodt}, \citenamefont {Song}, \citenamefont {Heindel},\ and\ \citenamefont {Reitzenstein}}]{Bremer:2020}%
  \BibitemOpen
  \bibfield  {author} {\bibinfo {author} {\bibfnamefont {L.}~\bibnamefont {Bremer}}, \bibinfo {author} {\bibfnamefont {S.}~\bibnamefont {Fischbach}}, \bibinfo {author} {\bibfnamefont {S.-I.}\ \bibnamefont {Park}}, \bibinfo {author} {\bibfnamefont {S.}~\bibnamefont {Rodt}}, \bibinfo {author} {\bibfnamefont {J.-D.}\ \bibnamefont {Song}}, \bibinfo {author} {\bibfnamefont {T.}~\bibnamefont {Heindel}},\ and\ \bibinfo {author} {\bibfnamefont {S.}~\bibnamefont {Reitzenstein}},\ }\bibfield  {title} {\bibinfo {title} {Cesium-vapor-based delay of single photons emitted by deterministically fabricated quantum dot microlenses},\ }\href {https://doi.org/https://doi.org/10.1002/qute.201900071} {\bibfield  {journal} {\bibinfo  {journal} {Advanced Quantum Technologies}\ }\textbf {\bibinfo {volume} {3}},\ \bibinfo {pages} {1900071} (\bibinfo {year} {2020})},\ \Eprint {https://arxiv.org/abs/https://onlinelibrary.wiley.com/doi/pdf/10.1002/qute.201900071} {https://onlinelibrary.wiley.com/doi/pdf/10.1002/qute.201900071}
  \BibitemShut {NoStop}%
\bibitem [{\citenamefont {Vural}\ \emph {et~al.}(2021)\citenamefont {Vural}, \citenamefont {Seyfferle}, \citenamefont {Gerhardt}, \citenamefont {Jetter}, \citenamefont {Portalupi},\ and\ \citenamefont {Michler}}]{Vural.2021}%
  \BibitemOpen
  \bibfield  {author} {\bibinfo {author} {\bibfnamefont {H.}~\bibnamefont {Vural}}, \bibinfo {author} {\bibfnamefont {S.}~\bibnamefont {Seyfferle}}, \bibinfo {author} {\bibfnamefont {I.}~\bibnamefont {Gerhardt}}, \bibinfo {author} {\bibfnamefont {M.}~\bibnamefont {Jetter}}, \bibinfo {author} {\bibfnamefont {S.~L.}\ \bibnamefont {Portalupi}},\ and\ \bibinfo {author} {\bibfnamefont {P.}~\bibnamefont {Michler}},\ }\bibfield  {title} {\bibinfo {title} {Delaying two-photon {F}ock states in hot cesium vapor using single photons generated on demand from a semiconductor quantum dot},\ }\href {https://doi.org/10.1103/PhysRevB.103.195304} {\bibfield  {journal} {\bibinfo  {journal} {Phys. Rev. B}\ }\textbf {\bibinfo {volume} {103}},\ \bibinfo {pages} {195304} (\bibinfo {year} {2021})}\BibitemShut {NoStop}%
\bibitem [{\citenamefont {Trotta}\ \emph {et~al.}(2016)\citenamefont {Trotta}, \citenamefont {Mart{\'i}n-S{\'a}nchez}, \citenamefont {Wildmann}, \citenamefont {Piredda}, \citenamefont {Reindl}, \citenamefont {Schimpf}, \citenamefont {Zallo}, \citenamefont {Stroj}, \citenamefont {Edlinger},\ and\ \citenamefont {Rastelli}}]{Trotta.2016}%
  \BibitemOpen
  \bibfield  {author} {\bibinfo {author} {\bibfnamefont {R.}~\bibnamefont {Trotta}}, \bibinfo {author} {\bibfnamefont {J.}~\bibnamefont {Mart{\'i}n-S{\'a}nchez}}, \bibinfo {author} {\bibfnamefont {J.~S.}\ \bibnamefont {Wildmann}}, \bibinfo {author} {\bibfnamefont {G.}~\bibnamefont {Piredda}}, \bibinfo {author} {\bibfnamefont {M.}~\bibnamefont {Reindl}}, \bibinfo {author} {\bibfnamefont {C.}~\bibnamefont {Schimpf}}, \bibinfo {author} {\bibfnamefont {E.}~\bibnamefont {Zallo}}, \bibinfo {author} {\bibfnamefont {S.}~\bibnamefont {Stroj}}, \bibinfo {author} {\bibfnamefont {J.}~\bibnamefont {Edlinger}},\ and\ \bibinfo {author} {\bibfnamefont {A.}~\bibnamefont {Rastelli}},\ }\bibfield  {title} {\bibinfo {title} {Wavelength-tunable sources of entangled photons interfaced with atomic vapours},\ }\href {https://doi.org/10.1038/ncomms10375} {\bibfield  {journal} {\bibinfo  {journal} {Nature Communications}\ }\textbf {\bibinfo {volume} {7}},\ \bibinfo {pages} {10375} (\bibinfo {year} {2016})}\BibitemShut {NoStop}%
\bibitem [{\citenamefont {Kaczmarek}\ \emph {et~al.}(2018)\citenamefont {Kaczmarek}, \citenamefont {Ledingham}, \citenamefont {Brecht}, \citenamefont {Thomas}, \citenamefont {Thekkadath}, \citenamefont {Lazo-Arjona}, \citenamefont {Munns}, \citenamefont {Poem}, \citenamefont {Feizpour}, \citenamefont {Saunders}, \citenamefont {Nunn},\ and\ \citenamefont {Walmsley}}]{Kaczmarek.2018}%
  \BibitemOpen
  \bibfield  {author} {\bibinfo {author} {\bibfnamefont {K.~T.}\ \bibnamefont {Kaczmarek}}, \bibinfo {author} {\bibfnamefont {P.~M.}\ \bibnamefont {Ledingham}}, \bibinfo {author} {\bibfnamefont {B.}~\bibnamefont {Brecht}}, \bibinfo {author} {\bibfnamefont {S.~E.}\ \bibnamefont {Thomas}}, \bibinfo {author} {\bibfnamefont {G.~S.}\ \bibnamefont {Thekkadath}}, \bibinfo {author} {\bibfnamefont {O.}~\bibnamefont {Lazo-Arjona}}, \bibinfo {author} {\bibfnamefont {J.~H.~D.}\ \bibnamefont {Munns}}, \bibinfo {author} {\bibfnamefont {E.}~\bibnamefont {Poem}}, \bibinfo {author} {\bibfnamefont {A.}~\bibnamefont {Feizpour}}, \bibinfo {author} {\bibfnamefont {D.~J.}\ \bibnamefont {Saunders}}, \bibinfo {author} {\bibfnamefont {J.}~\bibnamefont {Nunn}},\ and\ \bibinfo {author} {\bibfnamefont {I.~A.}\ \bibnamefont {Walmsley}},\ }\bibfield  {title} {\bibinfo {title} {High-speed noise-free optical quantum memory},\ }\href {https://doi.org/10.1103/PhysRevA.97.042316} {\bibfield  {journal} {\bibinfo  {journal} {Phys. Rev. A}\
  }\textbf {\bibinfo {volume} {97}},\ \bibinfo {pages} {042316} (\bibinfo {year} {2018})}\BibitemShut {NoStop}%
\bibitem [{\citenamefont {{Ran Finkelstein}}\ \emph {et~al.}(2018)\citenamefont {{Ran Finkelstein}}, \citenamefont {{Eilon Poem}}, \citenamefont {{Ohad Michel}}, \citenamefont {{Ohr Lahad}},\ and\ \citenamefont {{Ofer Firstenberg}}}]{RanFinkelstein.2018}%
  \BibitemOpen
  \bibfield  {author} {\bibinfo {author} {\bibnamefont {{Ran Finkelstein}}}, \bibinfo {author} {\bibnamefont {{Eilon Poem}}}, \bibinfo {author} {\bibnamefont {{Ohad Michel}}}, \bibinfo {author} {\bibnamefont {{Ohr Lahad}}},\ and\ \bibinfo {author} {\bibnamefont {{Ofer Firstenberg}}},\ }\bibfield  {title} {\bibinfo {title} {Fast, noise-free memory for photon synchronization at room temperature},\ }\href {https://doi.org/10.1126/sciadv.aap8598} {\bibfield  {journal} {\bibinfo  {journal} {Science advances}\ }\textbf {\bibinfo {volume} {4}},\ \bibinfo {pages} {eaap8598} (\bibinfo {year} {2018})}\BibitemShut {NoStop}%
\bibitem [{\citenamefont {Wolters}\ \emph {et~al.}(2017)\citenamefont {Wolters}, \citenamefont {Buser}, \citenamefont {Horsley}, \citenamefont {B\'eguin}, \citenamefont {J\"ockel}, \citenamefont {Jahn}, \citenamefont {Warburton},\ and\ \citenamefont {Treutlein}}]{Wolters.2017}%
  \BibitemOpen
  \bibfield  {author} {\bibinfo {author} {\bibfnamefont {J.}~\bibnamefont {Wolters}}, \bibinfo {author} {\bibfnamefont {G.}~\bibnamefont {Buser}}, \bibinfo {author} {\bibfnamefont {A.}~\bibnamefont {Horsley}}, \bibinfo {author} {\bibfnamefont {L.}~\bibnamefont {B\'eguin}}, \bibinfo {author} {\bibfnamefont {A.}~\bibnamefont {J\"ockel}}, \bibinfo {author} {\bibfnamefont {J.-P.}\ \bibnamefont {Jahn}}, \bibinfo {author} {\bibfnamefont {R.~J.}\ \bibnamefont {Warburton}},\ and\ \bibinfo {author} {\bibfnamefont {P.}~\bibnamefont {Treutlein}},\ }\bibfield  {title} {\bibinfo {title} {Simple atomic quantum memory suitable for semiconductor quantum dot single photons},\ }\href {https://doi.org/10.1103/PhysRevLett.119.060502} {\bibfield  {journal} {\bibinfo  {journal} {Phys. Rev. Lett.}\ }\textbf {\bibinfo {volume} {119}},\ \bibinfo {pages} {060502} (\bibinfo {year} {2017})}\BibitemShut {NoStop}%
\bibitem [{\citenamefont {Thomas}\ \emph {et~al.}(2023)\citenamefont {Thomas}, \citenamefont {Sagona-Stophel}, \citenamefont {Schofield}, \citenamefont {Walmsley},\ and\ \citenamefont {Ledingham}}]{Thomas.2023}%
  \BibitemOpen
  \bibfield  {author} {\bibinfo {author} {\bibfnamefont {S.}~\bibnamefont {Thomas}}, \bibinfo {author} {\bibfnamefont {S.}~\bibnamefont {Sagona-Stophel}}, \bibinfo {author} {\bibfnamefont {Z.}~\bibnamefont {Schofield}}, \bibinfo {author} {\bibfnamefont {I.}~\bibnamefont {Walmsley}},\ and\ \bibinfo {author} {\bibfnamefont {P.}~\bibnamefont {Ledingham}},\ }\bibfield  {title} {\bibinfo {title} {Single-photon-compatible telecommunications-band quantum memory in a hot atomic gas},\ }\href {https://doi.org/10.1103/PhysRevApplied.19.L031005} {\bibfield  {journal} {\bibinfo  {journal} {Phys. Rev. Appl.}\ }\textbf {\bibinfo {volume} {19}},\ \bibinfo {pages} {L031005} (\bibinfo {year} {2023})}\BibitemShut {NoStop}%
\bibitem [{\citenamefont {Maa\ss{}}\ \emph {et~al.}(2024)\citenamefont {Maa\ss{}}, \citenamefont {Ewald}, \citenamefont {Barua}, \citenamefont {Reitzenstein},\ and\ \citenamefont {Wolters}}]{Maaß.2024}%
  \BibitemOpen
  \bibfield  {author} {\bibinfo {author} {\bibfnamefont {B.}~\bibnamefont {Maa\ss{}}}, \bibinfo {author} {\bibfnamefont {N.~V.}\ \bibnamefont {Ewald}}, \bibinfo {author} {\bibfnamefont {A.}~\bibnamefont {Barua}}, \bibinfo {author} {\bibfnamefont {S.}~\bibnamefont {Reitzenstein}},\ and\ \bibinfo {author} {\bibfnamefont {J.}~\bibnamefont {Wolters}},\ }\bibfield  {title} {\bibinfo {title} {Room-temperature ladder-type optical memory compatible with single photons from semiconductor quantum dots},\ }\href {https://doi.org/10.1103/PhysRevApplied.22.044050} {\bibfield  {journal} {\bibinfo  {journal} {Phys. Rev. Appl.}\ }\textbf {\bibinfo {volume} {22}},\ \bibinfo {pages} {044050} (\bibinfo {year} {2024})}\BibitemShut {NoStop}%
\bibitem [{\citenamefont {Thomas}\ \emph {et~al.}(2024)\citenamefont {Thomas}, \citenamefont {Wagner}, \citenamefont {Joos}, \citenamefont {Sittig}, \citenamefont {Nawrath}, \citenamefont {Burdekin}, \citenamefont {de~{Buy Wenniger}}, \citenamefont {Rasiah}, \citenamefont {Huber-Loyola}, \citenamefont {Sagona-Stophel}, \citenamefont {H{\"o}fling}, \citenamefont {Jetter}, \citenamefont {Michler}, \citenamefont {Walmsley}, \citenamefont {Portalupi},\ and\ \citenamefont {Ledingham}}]{Thomas.2024}%
  \BibitemOpen
  \bibfield  {author} {\bibinfo {author} {\bibfnamefont {S.~E.}\ \bibnamefont {Thomas}}, \bibinfo {author} {\bibfnamefont {L.}~\bibnamefont {Wagner}}, \bibinfo {author} {\bibfnamefont {R.}~\bibnamefont {Joos}}, \bibinfo {author} {\bibfnamefont {R.}~\bibnamefont {Sittig}}, \bibinfo {author} {\bibfnamefont {C.}~\bibnamefont {Nawrath}}, \bibinfo {author} {\bibfnamefont {P.}~\bibnamefont {Burdekin}}, \bibinfo {author} {\bibfnamefont {I.~M.}\ \bibnamefont {de~{Buy Wenniger}}}, \bibinfo {author} {\bibfnamefont {M.~J.}\ \bibnamefont {Rasiah}}, \bibinfo {author} {\bibfnamefont {T.}~\bibnamefont {Huber-Loyola}}, \bibinfo {author} {\bibfnamefont {S.}~\bibnamefont {Sagona-Stophel}}, \bibinfo {author} {\bibfnamefont {S.}~\bibnamefont {H{\"o}fling}}, \bibinfo {author} {\bibfnamefont {M.}~\bibnamefont {Jetter}}, \bibinfo {author} {\bibfnamefont {P.}~\bibnamefont {Michler}}, \bibinfo {author} {\bibfnamefont {I.~A.}\ \bibnamefont {Walmsley}}, \bibinfo {author} {\bibfnamefont {S.~L.}\ \bibnamefont {Portalupi}},\ and\ \bibinfo
  {author} {\bibfnamefont {P.~M.}\ \bibnamefont {Ledingham}},\ }\bibfield  {title} {\bibinfo {title} {Deterministic storage and retrieval of telecom light from a quantum dot single-photon source interfaced with an atomic quantum memory},\ }\href {https://doi.org/10.1126/sciadv.adi7346} {\bibfield  {journal} {\bibinfo  {journal} {Science advances}\ }\textbf {\bibinfo {volume} {10}},\ \bibinfo {pages} {eadi7346} (\bibinfo {year} {2024})}\BibitemShut {NoStop}%
\bibitem [{\citenamefont {Rodt}\ and\ \citenamefont {Reitzenstein}(2021)}]{Rodt_2021}%
  \BibitemOpen
  \bibfield  {author} {\bibinfo {author} {\bibfnamefont {S.}~\bibnamefont {Rodt}}\ and\ \bibinfo {author} {\bibfnamefont {S.}~\bibnamefont {Reitzenstein}},\ }\bibfield  {title} {\bibinfo {title} {High-performance deterministic in situ electron-beam lithography enabled by cathodoluminescence spectroscopy},\ }\href {https://doi.org/10.1088/2632-959X/abed3c} {\bibfield  {journal} {\bibinfo  {journal} {Nano Express}\ }\textbf {\bibinfo {volume} {2}},\ \bibinfo {pages} {014007} (\bibinfo {year} {2021})}\BibitemShut {NoStop}%
\bibitem [{\citenamefont {Rodt}\ \emph {et~al.}(2020)\citenamefont {Rodt}, \citenamefont {Reitzenstein},\ and\ \citenamefont {Heindel}}]{Rodt_2020}%
  \BibitemOpen
  \bibfield  {author} {\bibinfo {author} {\bibfnamefont {S.}~\bibnamefont {Rodt}}, \bibinfo {author} {\bibfnamefont {S.}~\bibnamefont {Reitzenstein}},\ and\ \bibinfo {author} {\bibfnamefont {T.}~\bibnamefont {Heindel}},\ }\bibfield  {title} {\bibinfo {title} {Deterministically fabricated solid-state quantum-light sources},\ }\href {https://doi.org/10.1088/1361-648X/ab5e15} {\bibfield  {journal} {\bibinfo  {journal} {Journal of Physics: Condensed Matter}\ }\textbf {\bibinfo {volume} {32}},\ \bibinfo {pages} {153003} (\bibinfo {year} {2020})}\BibitemShut {NoStop}%
\bibitem [{\citenamefont {Rickert}\ \emph {et~al.}(2019)\citenamefont {Rickert}, \citenamefont {Kupko}, \citenamefont {Rodt}, \citenamefont {Reitzenstein},\ and\ \citenamefont {Heindel}}]{Rickert:19}%
  \BibitemOpen
  \bibfield  {author} {\bibinfo {author} {\bibfnamefont {L.}~\bibnamefont {Rickert}}, \bibinfo {author} {\bibfnamefont {T.}~\bibnamefont {Kupko}}, \bibinfo {author} {\bibfnamefont {S.}~\bibnamefont {Rodt}}, \bibinfo {author} {\bibfnamefont {S.}~\bibnamefont {Reitzenstein}},\ and\ \bibinfo {author} {\bibfnamefont {T.}~\bibnamefont {Heindel}},\ }\bibfield  {title} {\bibinfo {title} {Optimized designs for telecom-wavelength quantum light sources based on hybrid circular {B}ragg gratings},\ }\href {https://doi.org/10.1364/OE.27.036824} {\bibfield  {journal} {\bibinfo  {journal} {Opt. Express}\ }\textbf {\bibinfo {volume} {27}},\ \bibinfo {pages} {36824} (\bibinfo {year} {2019})}\BibitemShut {NoStop}%
\bibitem [{\citenamefont {Yao}\ \emph {et~al.}(2018)\citenamefont {Yao}, \citenamefont {Rongbin~Su}, \citenamefont {Zhuojun~Liu},\ and\ \citenamefont {Liu}}]{YaoBCBG}%
  \BibitemOpen
  \bibfield  {author} {\bibinfo {author} {\bibfnamefont {B.}~\bibnamefont {Yao}}, \bibinfo {author} {\bibfnamefont {Y.~W.}\ \bibnamefont {Rongbin~Su}}, \bibinfo {author} {\bibfnamefont {T.~Z.}\ \bibnamefont {Zhuojun~Liu}},\ and\ \bibinfo {author} {\bibfnamefont {J.}~\bibnamefont {Liu}},\ }\bibfield  {title} {\bibinfo {title} {Design for hybrid circular bragg gratings for a highly efficient quantum dot single-photon source},\ }\href@noop {} {\bibfield  {journal} {\bibinfo  {journal} {Journal of the Korean Physical Society}\ }\textbf {\bibinfo {volume} {73}} (\bibinfo {year} {2018})}\BibitemShut {NoStop}%
\bibitem [{\citenamefont {Pohl}(2020)}]{MBEgrowth}%
  \BibitemOpen
  \bibfield  {author} {\bibinfo {author} {\bibfnamefont {U.~W.}\ \bibnamefont {Pohl}},\ }\href@noop {} {\emph {\bibinfo {title} {Epitaxy of Semiconductors}}}\ (\bibinfo  {publisher} {Springer Nature},\ \bibinfo {year} {2020})\BibitemShut {NoStop}%
\bibitem [{\citenamefont {Wang}\ \emph {et~al.}(2019)\citenamefont {Wang}, \citenamefont {Hu}, \citenamefont {Chung}, \citenamefont {Qin}, \citenamefont {Yang}, \citenamefont {Li}, \citenamefont {Liu}, \citenamefont {Zhong}, \citenamefont {He}, \citenamefont {Ding}, \citenamefont {Deng}, \citenamefont {Dai}, \citenamefont {Huo}, \citenamefont {H\"ofling}, \citenamefont {Lu},\ and\ \citenamefont {Pan}}]{Wang2019}%
  \BibitemOpen
  \bibfield  {author} {\bibinfo {author} {\bibfnamefont {H.}~\bibnamefont {Wang}}, \bibinfo {author} {\bibfnamefont {H.}~\bibnamefont {Hu}}, \bibinfo {author} {\bibfnamefont {T.-H.}\ \bibnamefont {Chung}}, \bibinfo {author} {\bibfnamefont {J.}~\bibnamefont {Qin}}, \bibinfo {author} {\bibfnamefont {X.}~\bibnamefont {Yang}}, \bibinfo {author} {\bibfnamefont {J.-P.}\ \bibnamefont {Li}}, \bibinfo {author} {\bibfnamefont {R.-Z.}\ \bibnamefont {Liu}}, \bibinfo {author} {\bibfnamefont {H.-S.}\ \bibnamefont {Zhong}}, \bibinfo {author} {\bibfnamefont {Y.-M.}\ \bibnamefont {He}}, \bibinfo {author} {\bibfnamefont {X.}~\bibnamefont {Ding}}, \bibinfo {author} {\bibfnamefont {Y.-H.}\ \bibnamefont {Deng}}, \bibinfo {author} {\bibfnamefont {Q.}~\bibnamefont {Dai}}, \bibinfo {author} {\bibfnamefont {Y.-H.}\ \bibnamefont {Huo}}, \bibinfo {author} {\bibfnamefont {S.}~\bibnamefont {H\"ofling}}, \bibinfo {author} {\bibfnamefont {C.-Y.}\ \bibnamefont {Lu}},\ and\ \bibinfo {author} {\bibfnamefont {J.-W.}\ \bibnamefont {Pan}},\
  }\bibfield  {title} {\bibinfo {title} {On-demand semiconductor source of entangled photons which simultaneously has high fidelity, efficiency, and indistinguishability},\ }\href {https://doi.org/10.1103/PhysRevLett.122.113602} {\bibfield  {journal} {\bibinfo  {journal} {Phys. Rev. Lett.}\ }\textbf {\bibinfo {volume} {122}},\ \bibinfo {pages} {113602} (\bibinfo {year} {2019})}\BibitemShut {NoStop}%
\bibitem [{\citenamefont {Liu}\ \emph {et~al.}(2019)\citenamefont {Liu}, \citenamefont {Su}, \citenamefont {Wei}, \citenamefont {Yao}, \citenamefont {da~Silva}, \citenamefont {Yu}, \citenamefont {Iles-Smith}, \citenamefont {Srinivasan}, \citenamefont {Rastelli}, \citenamefont {Li} \emph {et~al.}}]{Jinliu2019CBG}%
  \BibitemOpen
  \bibfield  {author} {\bibinfo {author} {\bibfnamefont {J.}~\bibnamefont {Liu}}, \bibinfo {author} {\bibfnamefont {R.}~\bibnamefont {Su}}, \bibinfo {author} {\bibfnamefont {Y.}~\bibnamefont {Wei}}, \bibinfo {author} {\bibfnamefont {B.}~\bibnamefont {Yao}}, \bibinfo {author} {\bibfnamefont {S.~F.~C.}\ \bibnamefont {da~Silva}}, \bibinfo {author} {\bibfnamefont {Y.}~\bibnamefont {Yu}}, \bibinfo {author} {\bibfnamefont {J.}~\bibnamefont {Iles-Smith}}, \bibinfo {author} {\bibfnamefont {K.}~\bibnamefont {Srinivasan}}, \bibinfo {author} {\bibfnamefont {A.}~\bibnamefont {Rastelli}}, \bibinfo {author} {\bibfnamefont {J.}~\bibnamefont {Li}}, \emph {et~al.},\ }\bibfield  {title} {\bibinfo {title} {A solid-state source of strongly entangled photon pairs with high brightness and indistinguishability},\ }\href@noop {} {\bibfield  {journal} {\bibinfo  {journal} {Nature Nanotechnology}\ }\textbf {\bibinfo {volume} {14}} (\bibinfo {year} {2019})}\BibitemShut {NoStop}%
\bibitem [{\citenamefont {Nawrath}\ \emph {et~al.}(2023)\citenamefont {Nawrath}, \citenamefont {Joos}, \citenamefont {Kolatschek}, \citenamefont {Bauer}, \citenamefont {Pruy}, \citenamefont {Hornung}, \citenamefont {Fischer}, \citenamefont {Huang}, \citenamefont {Vijayan}, \citenamefont {Sittig}, \citenamefont {Jetter}, \citenamefont {Portalupi},\ and\ \citenamefont {Michler}}]{Nawrath_Michler}%
  \BibitemOpen
  \bibfield  {author} {\bibinfo {author} {\bibfnamefont {C.}~\bibnamefont {Nawrath}}, \bibinfo {author} {\bibfnamefont {R.}~\bibnamefont {Joos}}, \bibinfo {author} {\bibfnamefont {S.}~\bibnamefont {Kolatschek}}, \bibinfo {author} {\bibfnamefont {S.}~\bibnamefont {Bauer}}, \bibinfo {author} {\bibfnamefont {P.}~\bibnamefont {Pruy}}, \bibinfo {author} {\bibfnamefont {F.}~\bibnamefont {Hornung}}, \bibinfo {author} {\bibfnamefont {J.}~\bibnamefont {Fischer}}, \bibinfo {author} {\bibfnamefont {J.}~\bibnamefont {Huang}}, \bibinfo {author} {\bibfnamefont {P.}~\bibnamefont {Vijayan}}, \bibinfo {author} {\bibfnamefont {R.}~\bibnamefont {Sittig}}, \bibinfo {author} {\bibfnamefont {M.}~\bibnamefont {Jetter}}, \bibinfo {author} {\bibfnamefont {S.~L.}\ \bibnamefont {Portalupi}},\ and\ \bibinfo {author} {\bibfnamefont {P.}~\bibnamefont {Michler}},\ }\bibfield  {title} {\bibinfo {title} {Bright source of purcell-enhanced, triggered, single photons in the telecom c-band},\ }\href
  {https://doi.org/https://doi.org/10.1002/qute.202300111} {\bibfield  {journal} {\bibinfo  {journal} {Advanced Quantum Technologies}\ }\textbf {\bibinfo {volume} {6}},\ \bibinfo {pages} {2300111} (\bibinfo {year} {2023})},\ \Eprint {https://arxiv.org/abs/https://onlinelibrary.wiley.com/doi/pdf/10.1002/qute.202300111} {https://onlinelibrary.wiley.com/doi/pdf/10.1002/qute.202300111} \BibitemShut {NoStop}%
\bibitem [{\citenamefont {Shih}\ \emph {et~al.}(2023)\citenamefont {Shih}, \citenamefont {Rodt},\ and\ \citenamefont {Reitzenstein}}]{Shih:23}%
  \BibitemOpen
  \bibfield  {author} {\bibinfo {author} {\bibfnamefont {C.-W.}\ \bibnamefont {Shih}}, \bibinfo {author} {\bibfnamefont {S.}~\bibnamefont {Rodt}},\ and\ \bibinfo {author} {\bibfnamefont {S.}~\bibnamefont {Reitzenstein}},\ }\bibfield  {title} {\bibinfo {title} {Universal design method for bright quantum light sources based on circular bragg grating cavities},\ }\href {https://doi.org/10.1364/OE.501495} {\bibfield  {journal} {\bibinfo  {journal} {Opt. Express}\ }\textbf {\bibinfo {volume} {31}},\ \bibinfo {pages} {35552} (\bibinfo {year} {2023})}\BibitemShut {NoStop}%
\bibitem [{\citenamefont {Madigawa}\ \emph {et~al.}(2024{\natexlab{a}})\citenamefont {Madigawa}, \citenamefont {Donges}, \citenamefont {Gaál}, \citenamefont {Li}, \citenamefont {Jacobsen}, \citenamefont {Liu}, \citenamefont {Dai}, \citenamefont {Su}, \citenamefont {Shang}, \citenamefont {Ni}, \citenamefont {Schall}, \citenamefont {Rodt}, \citenamefont {Niu}, \citenamefont {Gregersen}, \citenamefont {Reitzenstein},\ and\ \citenamefont {Munkhbat}}]{donges2024}%
  \BibitemOpen
  \bibfield  {author} {\bibinfo {author} {\bibfnamefont {A.~A.}\ \bibnamefont {Madigawa}}, \bibinfo {author} {\bibfnamefont {J.~N.}\ \bibnamefont {Donges}}, \bibinfo {author} {\bibfnamefont {B.}~\bibnamefont {Gaál}}, \bibinfo {author} {\bibfnamefont {S.}~\bibnamefont {Li}}, \bibinfo {author} {\bibfnamefont {M.~A.}\ \bibnamefont {Jacobsen}}, \bibinfo {author} {\bibfnamefont {H.}~\bibnamefont {Liu}}, \bibinfo {author} {\bibfnamefont {D.}~\bibnamefont {Dai}}, \bibinfo {author} {\bibfnamefont {X.}~\bibnamefont {Su}}, \bibinfo {author} {\bibfnamefont {X.}~\bibnamefont {Shang}}, \bibinfo {author} {\bibfnamefont {H.}~\bibnamefont {Ni}}, \bibinfo {author} {\bibfnamefont {J.}~\bibnamefont {Schall}}, \bibinfo {author} {\bibfnamefont {S.}~\bibnamefont {Rodt}}, \bibinfo {author} {\bibfnamefont {Z.}~\bibnamefont {Niu}}, \bibinfo {author} {\bibfnamefont {N.}~\bibnamefont {Gregersen}}, \bibinfo {author} {\bibfnamefont {S.}~\bibnamefont {Reitzenstein}},\ and\ \bibinfo {author} {\bibfnamefont {B.}~\bibnamefont {Munkhbat}},\
  }\bibfield  {title} {\bibinfo {title} {Assessing the alignment accuracy of state-of-the-art deterministic fabrication methods for single quantum dot devices},\ }\href {https://doi.org/10.1021/acsphotonics.3c01368} {\bibfield  {journal} {\bibinfo  {journal} {ACS Photonics}\ }\textbf {\bibinfo {volume} {11}},\ \bibinfo {pages} {1012} (\bibinfo {year} {2024}{\natexlab{a}})},\ \Eprint {https://arxiv.org/abs/https://doi.org/10.1021/acsphotonics.3c01368} {https://doi.org/10.1021/acsphotonics.3c01368} \BibitemShut {NoStop}%
\bibitem [{\citenamefont {Burger}\ \emph {et~al.}(2015)\citenamefont {Burger}, \citenamefont {Zschiedrich}, \citenamefont {Pomplun}, \citenamefont {Herrmann},\ and\ \citenamefont {Schmidt}}]{Burger_JCM}%
  \BibitemOpen
  \bibfield  {author} {\bibinfo {author} {\bibfnamefont {S.}~\bibnamefont {Burger}}, \bibinfo {author} {\bibfnamefont {L.}~\bibnamefont {Zschiedrich}}, \bibinfo {author} {\bibfnamefont {J.}~\bibnamefont {Pomplun}}, \bibinfo {author} {\bibfnamefont {S.}~\bibnamefont {Herrmann}},\ and\ \bibinfo {author} {\bibfnamefont {F.}~\bibnamefont {Schmidt}},\ }\bibfield  {title} {\bibinfo {title} {{Hp-finite element method for simulating light scattering from complex 3D structures}},\ }in\ \href {https://doi.org/10.1117/12.2085795} {\emph {\bibinfo {booktitle} {Metrology, Inspection, and Process Control for Microlithography XXIX}}},\ Vol.\ \bibinfo {volume} {9424},\ \bibinfo {editor} {edited by\ \bibinfo {editor} {\bibfnamefont {J.~P.}\ \bibnamefont {Cain}}\ and\ \bibinfo {editor} {\bibfnamefont {M.~I.}\ \bibnamefont {Sanchez}}},\ \bibinfo {organization} {International Society for Optics and Photonics}\ (\bibinfo  {publisher} {SPIE},\ \bibinfo {year} {2015})\ p.\ \bibinfo {pages} {94240Z}\BibitemShut {NoStop}%
\bibitem [{\citenamefont {Schneider}(2024)}]{JCMWebsite}%
  \BibitemOpen
  \bibfield  {author} {\bibinfo {author} {\bibfnamefont {P.-I.}\ \bibnamefont {Schneider}},\ }\href {https://jcmwave.com/jcmsuite} {\bibinfo {title} {Jcmsuite}} (\bibinfo {year} {14.08.2024})\BibitemShut {NoStop}%
\bibitem [{\citenamefont {Hanbury~Brown}\ and\ \citenamefont {Twiss}(1956)}]{HBT}%
  \BibitemOpen
  \bibfield  {author} {\bibinfo {author} {\bibfnamefont {R.}~\bibnamefont {Hanbury~Brown}}\ and\ \bibinfo {author} {\bibfnamefont {R.}~\bibnamefont {Twiss}},\ }\bibfield  {title} {\bibinfo {title} {The question of correlation between photons in coherent light rays},\ }\href@noop {} {\bibfield  {journal} {\bibinfo  {journal} {Nature}\ }\textbf {\bibinfo {volume} {178}} (\bibinfo {year} {1956})}\BibitemShut {NoStop}%
\bibitem [{\citenamefont {Ding}\ \emph {et~al.}(2016)\citenamefont {Ding}, \citenamefont {He}, \citenamefont {Duan}, \citenamefont {Gregersen}, \citenamefont {Chen}, \citenamefont {Unsleber}, \citenamefont {Maier}, \citenamefont {Schneider}, \citenamefont {Kamp}, \citenamefont {H\"ofling}, \citenamefont {Lu},\ and\ \citenamefont {Pan}}]{Jianweipan2016}%
  \BibitemOpen
  \bibfield  {author} {\bibinfo {author} {\bibfnamefont {X.}~\bibnamefont {Ding}}, \bibinfo {author} {\bibfnamefont {Y.}~\bibnamefont {He}}, \bibinfo {author} {\bibfnamefont {Z.-C.}\ \bibnamefont {Duan}}, \bibinfo {author} {\bibfnamefont {N.}~\bibnamefont {Gregersen}}, \bibinfo {author} {\bibfnamefont {M.-C.}\ \bibnamefont {Chen}}, \bibinfo {author} {\bibfnamefont {S.}~\bibnamefont {Unsleber}}, \bibinfo {author} {\bibfnamefont {S.}~\bibnamefont {Maier}}, \bibinfo {author} {\bibfnamefont {C.}~\bibnamefont {Schneider}}, \bibinfo {author} {\bibfnamefont {M.}~\bibnamefont {Kamp}}, \bibinfo {author} {\bibfnamefont {S.}~\bibnamefont {H\"ofling}}, \bibinfo {author} {\bibfnamefont {C.-Y.}\ \bibnamefont {Lu}},\ and\ \bibinfo {author} {\bibfnamefont {J.-W.}\ \bibnamefont {Pan}},\ }\bibfield  {title} {\bibinfo {title} {On-demand single photons with high extraction efficiency and near-unity indistinguishability from a resonantly driven quantum dot in a micropillar},\ }\href
  {https://doi.org/10.1103/PhysRevLett.116.020401} {\bibfield  {journal} {\bibinfo  {journal} {Phys. Rev. Lett.}\ }\textbf {\bibinfo {volume} {116}},\ \bibinfo {pages} {020401} (\bibinfo {year} {2016})}\BibitemShut {NoStop}%
\bibitem [{\citenamefont {Santori}\ \emph {et~al.}(2004)\citenamefont {Santori}, \citenamefont {Fattal}, \citenamefont {Vuckovic}, \citenamefont {Solomon},\ and\ \citenamefont {Yamamoto}}]{Santori_2004}%
  \BibitemOpen
  \bibfield  {author} {\bibinfo {author} {\bibfnamefont {C.}~\bibnamefont {Santori}}, \bibinfo {author} {\bibfnamefont {D.}~\bibnamefont {Fattal}}, \bibinfo {author} {\bibfnamefont {J.}~\bibnamefont {Vuckovic}}, \bibinfo {author} {\bibfnamefont {G.~S.}\ \bibnamefont {Solomon}},\ and\ \bibinfo {author} {\bibfnamefont {Y.}~\bibnamefont {Yamamoto}},\ }\bibfield  {title} {\bibinfo {title} {Single-photon generation with inas quantum dots},\ }\href {https://doi.org/10.1088/1367-2630/6/1/089} {\bibfield  {journal} {\bibinfo  {journal} {New Journal of Physics}\ }\textbf {\bibinfo {volume} {6}},\ \bibinfo {pages} {89} (\bibinfo {year} {2004})}\BibitemShut {NoStop}%
\bibitem [{\citenamefont {Gao}\ \emph {et~al.}(2019)\citenamefont {Gao}, \citenamefont {Lazo-Arjona}, \citenamefont {Brecht}, \citenamefont {Kaczmarek}, \citenamefont {Thomas}, \citenamefont {Nunn}, \citenamefont {Ledingham}, \citenamefont {Saunders},\ and\ \citenamefont {Walmsley}}]{Gao.2019}%
  \BibitemOpen
  \bibfield  {author} {\bibinfo {author} {\bibfnamefont {S.}~\bibnamefont {Gao}}, \bibinfo {author} {\bibfnamefont {O.}~\bibnamefont {Lazo-Arjona}}, \bibinfo {author} {\bibfnamefont {B.}~\bibnamefont {Brecht}}, \bibinfo {author} {\bibfnamefont {K.~T.}\ \bibnamefont {Kaczmarek}}, \bibinfo {author} {\bibfnamefont {S.~E.}\ \bibnamefont {Thomas}}, \bibinfo {author} {\bibfnamefont {J.}~\bibnamefont {Nunn}}, \bibinfo {author} {\bibfnamefont {P.~M.}\ \bibnamefont {Ledingham}}, \bibinfo {author} {\bibfnamefont {D.~J.}\ \bibnamefont {Saunders}},\ and\ \bibinfo {author} {\bibfnamefont {I.~A.}\ \bibnamefont {Walmsley}},\ }\bibfield  {title} {\bibinfo {title} {Optimal coherent filtering for single noisy photons},\ }\href {https://doi.org/10.1103/PhysRevLett.123.213604} {\bibfield  {journal} {\bibinfo  {journal} {Physical Review Letters}\ }\textbf {\bibinfo {volume} {123}},\ \bibinfo {pages} {213604} (\bibinfo {year} {2019})}\BibitemShut {NoStop}%
\bibitem [{\citenamefont {Zhai}\ \emph {et~al.}(2020{\natexlab{b}})\citenamefont {Zhai}, \citenamefont {Löbl}, \citenamefont {Nguyen}, \citenamefont {Ritzmann}, \citenamefont {Javadi}, \citenamefont {Spinnler}, \citenamefont {Wieck}, \citenamefont {Ludwig},\ and\ \citenamefont {Warburton}}]{Richard2020}%
  \BibitemOpen
  \bibfield  {author} {\bibinfo {author} {\bibfnamefont {L.}~\bibnamefont {Zhai}}, \bibinfo {author} {\bibfnamefont {M.~C.}\ \bibnamefont {Löbl}}, \bibinfo {author} {\bibfnamefont {G.~N.}\ \bibnamefont {Nguyen}}, \bibinfo {author} {\bibfnamefont {J.}~\bibnamefont {Ritzmann}}, \bibinfo {author} {\bibfnamefont {A.}~\bibnamefont {Javadi}}, \bibinfo {author} {\bibfnamefont {C.}~\bibnamefont {Spinnler}}, \bibinfo {author} {\bibfnamefont {A.~D.}\ \bibnamefont {Wieck}}, \bibinfo {author} {\bibfnamefont {A.}~\bibnamefont {Ludwig}},\ and\ \bibinfo {author} {\bibfnamefont {R.~J.}\ \bibnamefont {Warburton}},\ }\bibfield  {title} {\bibinfo {title} {Low-noise {G}a{A}s quantum dots for quantum photonics},\ }\bibfield  {journal} {\bibinfo  {journal} {Nature Communications}\ }\textbf {\bibinfo {volume} {11}},\ \href {https://doi.org/10.1038/s41467-020-18625-z} {10.1038/s41467-020-18625-z} (\bibinfo {year} {2020}{\natexlab{b}})\BibitemShut {NoStop}%
\bibitem [{\citenamefont {Moczala-Dusanowska}\ \emph {et~al.}(2020)\citenamefont {Moczala-Dusanowska}, \citenamefont {Dusanowski}, \citenamefont {Iff}, \citenamefont {Huber}, \citenamefont {Kuhn}, \citenamefont {Czyszanowski}, \citenamefont {Schneider},\ and\ \citenamefont {Höfling}}]{Magdelena2020}%
  \BibitemOpen
  \bibfield  {author} {\bibinfo {author} {\bibfnamefont {M.}~\bibnamefont {Moczala-Dusanowska}}, \bibinfo {author} {\bibfnamefont {L.}~\bibnamefont {Dusanowski}}, \bibinfo {author} {\bibfnamefont {O.}~\bibnamefont {Iff}}, \bibinfo {author} {\bibfnamefont {T.}~\bibnamefont {Huber}}, \bibinfo {author} {\bibfnamefont {S.}~\bibnamefont {Kuhn}}, \bibinfo {author} {\bibfnamefont {T.}~\bibnamefont {Czyszanowski}}, \bibinfo {author} {\bibfnamefont {C.}~\bibnamefont {Schneider}},\ and\ \bibinfo {author} {\bibfnamefont {S.}~\bibnamefont {Höfling}},\ }\bibfield  {title} {\bibinfo {title} {Strain-tunable single-photon source based on a circular {B}ragg grating cavity with embedded quantum dots},\ }\href {https://doi.org/10.1021/acsphotonics.0c01465} {\bibfield  {journal} {\bibinfo  {journal} {ACS Photonics}\ }\textbf {\bibinfo {volume} {7}},\ \bibinfo {pages} {3474} (\bibinfo {year} {2020})},\ \Eprint {https://arxiv.org/abs/https://doi.org/10.1021/acsphotonics.0c01465} {https://doi.org/10.1021/acsphotonics.0c01465}
  \BibitemShut {NoStop}%
\bibitem [{\citenamefont {Wijitpatima}\ \emph {et~al.}(2024)\citenamefont {Wijitpatima}, \citenamefont {Auler}, \citenamefont {Mudi}, \citenamefont {Funk}, \citenamefont {Barua}, \citenamefont {Shrestha}, \citenamefont {Limame}, \citenamefont {Rodt}, \citenamefont {Reuter},\ and\ \citenamefont {Reitzenstein}}]{wijitpatima2024}%
  \BibitemOpen
  \bibfield  {author} {\bibinfo {author} {\bibfnamefont {S.}~\bibnamefont {Wijitpatima}}, \bibinfo {author} {\bibfnamefont {N.}~\bibnamefont {Auler}}, \bibinfo {author} {\bibfnamefont {P.}~\bibnamefont {Mudi}}, \bibinfo {author} {\bibfnamefont {T.}~\bibnamefont {Funk}}, \bibinfo {author} {\bibfnamefont {A.}~\bibnamefont {Barua}}, \bibinfo {author} {\bibfnamefont {B.}~\bibnamefont {Shrestha}}, \bibinfo {author} {\bibfnamefont {I.}~\bibnamefont {Limame}}, \bibinfo {author} {\bibfnamefont {S.}~\bibnamefont {Rodt}}, \bibinfo {author} {\bibfnamefont {D.}~\bibnamefont {Reuter}},\ and\ \bibinfo {author} {\bibfnamefont {S.}~\bibnamefont {Reitzenstein}},\ }\href {https://arxiv.org/abs/2406.08057} {\bibinfo {title} {Bright electrically contacted circular {B}ragg grating resonators with deterministically integrated quantum dots}} (\bibinfo {year} {2024}),\ \Eprint {https://arxiv.org/abs/2406.08057} {arXiv:2406.08057 [cond-mat.mes-hall]} \BibitemShut {NoStop}%
\bibitem [{\citenamefont {Main}\ \emph {et~al.}(2021)\citenamefont {Main}, \citenamefont {Hird}, \citenamefont {Gao}, \citenamefont {Oguz}, \citenamefont {Saunders}, \citenamefont {Walmsley},\ and\ \citenamefont {Ledingham}}]{Main.2021}%
  \BibitemOpen
  \bibfield  {author} {\bibinfo {author} {\bibfnamefont {D.}~\bibnamefont {Main}}, \bibinfo {author} {\bibfnamefont {T.~M.}\ \bibnamefont {Hird}}, \bibinfo {author} {\bibfnamefont {S.}~\bibnamefont {Gao}}, \bibinfo {author} {\bibfnamefont {E.}~\bibnamefont {Oguz}}, \bibinfo {author} {\bibfnamefont {D.~J.}\ \bibnamefont {Saunders}}, \bibinfo {author} {\bibfnamefont {I.~A.}\ \bibnamefont {Walmsley}},\ and\ \bibinfo {author} {\bibfnamefont {P.~M.}\ \bibnamefont {Ledingham}},\ }\bibfield  {title} {\bibinfo {title} {Preparing narrow velocity distributions for quantum memories in room-temperature alkali-metal vapors},\ }\href {https://doi.org/10.1103/PhysRevA.103.043105} {\bibfield  {journal} {\bibinfo  {journal} {Phys. Rev. A}\ }\textbf {\bibinfo {volume} {103}},\ \bibinfo {pages} {043105} (\bibinfo {year} {2021})}\BibitemShut {NoStop}%
\bibitem [{\citenamefont {Finkelstein}\ \emph {et~al.}(2021)\citenamefont {Finkelstein}, \citenamefont {Lahad}, \citenamefont {Cohen}, \citenamefont {Davidson}, \citenamefont {Kiriati}, \citenamefont {Poem},\ and\ \citenamefont {Firstenberg}}]{Finkelstein.2021}%
  \BibitemOpen
  \bibfield  {author} {\bibinfo {author} {\bibfnamefont {R.}~\bibnamefont {Finkelstein}}, \bibinfo {author} {\bibfnamefont {O.}~\bibnamefont {Lahad}}, \bibinfo {author} {\bibfnamefont {I.}~\bibnamefont {Cohen}}, \bibinfo {author} {\bibfnamefont {O.}~\bibnamefont {Davidson}}, \bibinfo {author} {\bibfnamefont {S.}~\bibnamefont {Kiriati}}, \bibinfo {author} {\bibfnamefont {E.}~\bibnamefont {Poem}},\ and\ \bibinfo {author} {\bibfnamefont {O.}~\bibnamefont {Firstenberg}},\ }\bibfield  {title} {\bibinfo {title} {Continuous protection of a collective state from inhomogeneous dephasing},\ }\href {https://doi.org/10.1103/PhysRevX.11.011008} {\bibfield  {journal} {\bibinfo  {journal} {Phys. Rev. X}\ }\textbf {\bibinfo {volume} {11}},\ \bibinfo {pages} {011008} (\bibinfo {year} {2021})}\BibitemShut {NoStop}%
\bibitem [{\citenamefont {{Roberto Mottola}}\ \emph {et~al.}(2020)\citenamefont {{Roberto Mottola}}, \citenamefont {{Gianni Buser}}, \citenamefont {{Chris M{\"u}ller}}, \citenamefont {{Tim Kroh}}, \citenamefont {{Andreas Ahlrichs}}, \citenamefont {{Sven Ramelow}}, \citenamefont {{Oliver Benson}}, \citenamefont {{Philipp Treutlein}},\ and\ \citenamefont {{Janik Wolters}}}]{Mottola.2020}%
  \BibitemOpen
  \bibfield  {author} {\bibinfo {author} {\bibnamefont {{Roberto Mottola}}}, \bibinfo {author} {\bibnamefont {{Gianni Buser}}}, \bibinfo {author} {\bibnamefont {{Chris M{\"u}ller}}}, \bibinfo {author} {\bibnamefont {{Tim Kroh}}}, \bibinfo {author} {\bibnamefont {{Andreas Ahlrichs}}}, \bibinfo {author} {\bibnamefont {{Sven Ramelow}}}, \bibinfo {author} {\bibnamefont {{Oliver Benson}}}, \bibinfo {author} {\bibnamefont {{Philipp Treutlein}}},\ and\ \bibinfo {author} {\bibnamefont {{Janik Wolters}}},\ }\bibfield  {title} {\bibinfo {title} {An efficient, tunable, and robust source of narrow-band photon pairs at the 87rb d1 line},\ }\href {https://doi.org/10.1364/OE.384081} {\bibfield  {journal} {\bibinfo  {journal} {Opt. Express}\ }\textbf {\bibinfo {volume} {28}},\ \bibinfo {pages} {3159} (\bibinfo {year} {2020})}\BibitemShut {NoStop}%
\bibitem [{\citenamefont {Buser}\ \emph {et~al.}(2022)\citenamefont {Buser}, \citenamefont {Mottola}, \citenamefont {Cotting}, \citenamefont {Wolters},\ and\ \citenamefont {Treutlein}}]{Buser.2022}%
  \BibitemOpen
  \bibfield  {author} {\bibinfo {author} {\bibfnamefont {G.}~\bibnamefont {Buser}}, \bibinfo {author} {\bibfnamefont {R.}~\bibnamefont {Mottola}}, \bibinfo {author} {\bibfnamefont {B.}~\bibnamefont {Cotting}}, \bibinfo {author} {\bibfnamefont {J.}~\bibnamefont {Wolters}},\ and\ \bibinfo {author} {\bibfnamefont {P.}~\bibnamefont {Treutlein}},\ }\bibfield  {title} {\bibinfo {title} {Single-photon storage in a ground-state vapor cell quantum memory},\ }\href {https://doi.org/10.1103/PRXQuantum.3.020349} {\bibfield  {journal} {\bibinfo  {journal} {PRX Quantum}\ }\textbf {\bibinfo {volume} {3}},\ \bibinfo {pages} {020349} (\bibinfo {year} {2022})}\BibitemShut {NoStop}%
\bibitem [{\citenamefont {Davidson}\ \emph {et~al.}(2023)\citenamefont {Davidson}, \citenamefont {Yogev}, \citenamefont {Poem},\ and\ \citenamefont {Firstenberg}}]{Davidson.2023}%
  \BibitemOpen
  \bibfield  {author} {\bibinfo {author} {\bibfnamefont {O.}~\bibnamefont {Davidson}}, \bibinfo {author} {\bibfnamefont {O.}~\bibnamefont {Yogev}}, \bibinfo {author} {\bibfnamefont {E.}~\bibnamefont {Poem}},\ and\ \bibinfo {author} {\bibfnamefont {O.}~\bibnamefont {Firstenberg}},\ }\bibfield  {title} {\bibinfo {title} {Single-photon synchronization with a room-temperature atomic quantum memory},\ }\href {https://doi.org/10.1103/PhysRevLett.131.033601} {\bibfield  {journal} {\bibinfo  {journal} {Phys. Rev. Lett.}\ }\textbf {\bibinfo {volume} {131}},\ \bibinfo {pages} {033601} (\bibinfo {year} {2023})}\BibitemShut {NoStop}%
\bibitem [{\citenamefont {Jenkins}(1990)}]{JenkinsOpticalconstants}%
  \BibitemOpen
  \bibfield  {author} {\bibinfo {author} {\bibfnamefont {D.~W.}\ \bibnamefont {Jenkins}},\ }\bibfield  {title} {\bibinfo {title} {{Optical constants of AlxGa1-xAs}},\ }\href {https://doi.org/10.1063/1.346621} {\bibfield  {journal} {\bibinfo  {journal} {Journal of Applied Physics}\ }\textbf {\bibinfo {volume} {68}},\ \bibinfo {pages} {1848} (\bibinfo {year} {1990})},\ \Eprint {https://arxiv.org/abs/https://pubs.aip.org/aip/jap/article-pdf/68/4/1848/18636134/1848\_1\_online.pdf} {https://pubs.aip.org/aip/jap/article-pdf/68/4/1848/18636134/1848\_1\_online.pdf} \BibitemShut {NoStop}%
\bibitem [{\citenamefont {Johnson}\ and\ \citenamefont {Christy}(1972)}]{JphnsonOpticalconsNobelMetal}%
  \BibitemOpen
  \bibfield  {author} {\bibinfo {author} {\bibfnamefont {P.~B.}\ \bibnamefont {Johnson}}\ and\ \bibinfo {author} {\bibfnamefont {R.~W.}\ \bibnamefont {Christy}},\ }\bibfield  {title} {\bibinfo {title} {Optical constants of the noble metals},\ }\href {https://doi.org/10.1103/PhysRevB.6.4370} {\bibfield  {journal} {\bibinfo  {journal} {Phys. Rev. B}\ }\textbf {\bibinfo {volume} {6}},\ \bibinfo {pages} {4370} (\bibinfo {year} {1972})}\BibitemShut {NoStop}%
\bibitem [{\citenamefont {Marple}(1964)}]{marpleRIGaAs}%
  \BibitemOpen
  \bibfield  {author} {\bibinfo {author} {\bibfnamefont {D.~T.~F.}\ \bibnamefont {Marple}},\ }\bibfield  {title} {\bibinfo {title} {{Refractive Index of GaAs}},\ }\href {https://doi.org/10.1063/1.1713601} {\bibfield  {journal} {\bibinfo  {journal} {Journal of Applied Physics}\ }\textbf {\bibinfo {volume} {35}},\ \bibinfo {pages} {1241} (\bibinfo {year} {1964})},\ \Eprint {https://arxiv.org/abs/https://pubs.aip.org/aip/jap/article-pdf/35/4/1241/18331384/1241\_1\_online.pdf} {https://pubs.aip.org/aip/jap/article-pdf/35/4/1241/18331384/1241\_1\_online.pdf} \BibitemShut {NoStop}%
\bibitem [{\citenamefont {de~Marcos}\ \emph {et~al.}(2016)\citenamefont {de~Marcos}, \citenamefont {Larruquert}, \citenamefont {M\'{e}ndez},\ and\ \citenamefont {Azn\'{a}rez}}]{RodriguezSiO2constants}%
  \BibitemOpen
  \bibfield  {author} {\bibinfo {author} {\bibfnamefont {L.~V.~R.}\ \bibnamefont {de~Marcos}}, \bibinfo {author} {\bibfnamefont {J.~I.}\ \bibnamefont {Larruquert}}, \bibinfo {author} {\bibfnamefont {J.~A.}\ \bibnamefont {M\'{e}ndez}},\ and\ \bibinfo {author} {\bibfnamefont {J.~A.}\ \bibnamefont {Azn\'{a}rez}},\ }\bibfield  {title} {\bibinfo {title} {Self-consistent optical constants of sio2 and ta2o5 films},\ }\href {https://doi.org/10.1364/OME.6.003622} {\bibfield  {journal} {\bibinfo  {journal} {Opt. Mater. Express}\ }\textbf {\bibinfo {volume} {6}},\ \bibinfo {pages} {3622} (\bibinfo {year} {2016})}\BibitemShut {NoStop}%
\bibitem [{\citenamefont {Gschrey}\ \emph {et~al.}(2015)\citenamefont {Gschrey}, \citenamefont {Schmidt}, \citenamefont {Schulze}, \citenamefont {Strittmatter}, \citenamefont {Rodt},\ and\ \citenamefont {Reitzenstein}}]{Gschrey.2015}%
  \BibitemOpen
  \bibfield  {author} {\bibinfo {author} {\bibfnamefont {M.}~\bibnamefont {Gschrey}}, \bibinfo {author} {\bibfnamefont {R.}~\bibnamefont {Schmidt}}, \bibinfo {author} {\bibfnamefont {J.-H.}\ \bibnamefont {Schulze}}, \bibinfo {author} {\bibfnamefont {A.}~\bibnamefont {Strittmatter}}, \bibinfo {author} {\bibfnamefont {S.}~\bibnamefont {Rodt}},\ and\ \bibinfo {author} {\bibfnamefont {S.}~\bibnamefont {Reitzenstein}},\ }\bibfield  {title} {\bibinfo {title} {{Resolution and alignment accuracy of low-temperature in situ electron beam lithography for nanophotonic device fabrication}},\ }\href {https://doi.org/10.1116/1.4914914} {\bibfield  {journal} {\bibinfo  {journal} {Journal of Vacuum Science \& Technology B}\ }\textbf {\bibinfo {volume} {33}},\ \bibinfo {pages} {021603} (\bibinfo {year} {2015})},\ \Eprint {https://arxiv.org/abs/https://pubs.aip.org/avs/jvb/article-pdf/doi/10.1116/1.4914914/13669261/021603\_1\_online.pdf}
  {https://pubs.aip.org/avs/jvb/article-pdf/doi/10.1116/1.4914914/13669261/021603\_1\_online.pdf} \BibitemShut {NoStop}%
\bibitem [{\citenamefont {Madigawa}\ \emph {et~al.}(2024{\natexlab{b}})\citenamefont {Madigawa}, \citenamefont {Donges}, \citenamefont {Gaál}, \citenamefont {Li}, \citenamefont {Jacobsen}, \citenamefont {Liu}, \citenamefont {Dai}, \citenamefont {Su}, \citenamefont {Shang}, \citenamefont {Ni}, \citenamefont {Schall}, \citenamefont {Rodt}, \citenamefont {Niu}, \citenamefont {Gregersen}, \citenamefont {Reitzenstein},\ and\ \citenamefont {Munkhbat}}]{Madigawa.2024}%
  \BibitemOpen
  \bibfield  {author} {\bibinfo {author} {\bibfnamefont {A.~A.}\ \bibnamefont {Madigawa}}, \bibinfo {author} {\bibfnamefont {J.~N.}\ \bibnamefont {Donges}}, \bibinfo {author} {\bibfnamefont {B.}~\bibnamefont {Gaál}}, \bibinfo {author} {\bibfnamefont {S.}~\bibnamefont {Li}}, \bibinfo {author} {\bibfnamefont {M.~A.}\ \bibnamefont {Jacobsen}}, \bibinfo {author} {\bibfnamefont {H.}~\bibnamefont {Liu}}, \bibinfo {author} {\bibfnamefont {D.}~\bibnamefont {Dai}}, \bibinfo {author} {\bibfnamefont {X.}~\bibnamefont {Su}}, \bibinfo {author} {\bibfnamefont {X.}~\bibnamefont {Shang}}, \bibinfo {author} {\bibfnamefont {H.}~\bibnamefont {Ni}}, \bibinfo {author} {\bibfnamefont {J.}~\bibnamefont {Schall}}, \bibinfo {author} {\bibfnamefont {S.}~\bibnamefont {Rodt}}, \bibinfo {author} {\bibfnamefont {Z.}~\bibnamefont {Niu}}, \bibinfo {author} {\bibfnamefont {N.}~\bibnamefont {Gregersen}}, \bibinfo {author} {\bibfnamefont {S.}~\bibnamefont {Reitzenstein}},\ and\ \bibinfo {author} {\bibfnamefont {B.}~\bibnamefont {Munkhbat}},\
  }\bibfield  {title} {\bibinfo {title} {Assessing the alignment accuracy of state-of-the-art deterministic fabrication methods for single quantum dot devices},\ }\href {https://doi.org/10.1021/acsphotonics.3c01368} {\bibfield  {journal} {\bibinfo  {journal} {ACS Photonics}\ }\textbf {\bibinfo {volume} {11}},\ \bibinfo {pages} {1012} (\bibinfo {year} {2024}{\natexlab{b}})},\ \Eprint {https://arxiv.org/abs/https://doi.org/10.1021/acsphotonics.3c01368} {https://doi.org/10.1021/acsphotonics.3c01368} \BibitemShut {NoStop}%
\bibitem [{\citenamefont {Cole}\ \emph {et~al.}(1992)\citenamefont {Cole}, \citenamefont {Salimian}, \citenamefont {Cooper}, \citenamefont {Lee},\ and\ \citenamefont {Dutta}}]{ICPRIE}%
  \BibitemOpen
  \bibfield  {author} {\bibinfo {author} {\bibfnamefont {M.~W.}\ \bibnamefont {Cole}}, \bibinfo {author} {\bibfnamefont {S.}~\bibnamefont {Salimian}}, \bibinfo {author} {\bibfnamefont {C.~B.}\ \bibnamefont {Cooper}}, \bibinfo {author} {\bibfnamefont {H.~S.}\ \bibnamefont {Lee}},\ and\ \bibinfo {author} {\bibfnamefont {M.}~\bibnamefont {Dutta}},\ }\bibfield  {title} {\bibinfo {title} {Reactive ion etching of gaas with sicl4: A residual damage and electrical investigation},\ }\href {https://doi.org/https://doi.org/10.1002/sca.4950140106} {\bibfield  {journal} {\bibinfo  {journal} {Scanning}\ }\textbf {\bibinfo {volume} {14}},\ \bibinfo {pages} {31} (\bibinfo {year} {1992})},\ \Eprint {https://arxiv.org/abs/https://onlinelibrary.wiley.com/doi/pdf/10.1002/sca.4950140106} {https://onlinelibrary.wiley.com/doi/pdf/10.1002/sca.4950140106} \BibitemShut {NoStop}%
\bibitem [{\citenamefont {Finley}\ \emph {et~al.}(2001)\citenamefont {Finley}, \citenamefont {Ashmore}, \citenamefont {Lema\^{\i}tre}, \citenamefont {Mowbray}, \citenamefont {Skolnick}, \citenamefont {Itskevich}, \citenamefont {Maksym}, \citenamefont {Hopkinson},\ and\ \citenamefont {Krauss}}]{Finley2001}%
  \BibitemOpen
  \bibfield  {author} {\bibinfo {author} {\bibfnamefont {J.~J.}\ \bibnamefont {Finley}}, \bibinfo {author} {\bibfnamefont {A.~D.}\ \bibnamefont {Ashmore}}, \bibinfo {author} {\bibfnamefont {A.}~\bibnamefont {Lema\^{\i}tre}}, \bibinfo {author} {\bibfnamefont {D.~J.}\ \bibnamefont {Mowbray}}, \bibinfo {author} {\bibfnamefont {M.~S.}\ \bibnamefont {Skolnick}}, \bibinfo {author} {\bibfnamefont {I.~E.}\ \bibnamefont {Itskevich}}, \bibinfo {author} {\bibfnamefont {P.~A.}\ \bibnamefont {Maksym}}, \bibinfo {author} {\bibfnamefont {M.}~\bibnamefont {Hopkinson}},\ and\ \bibinfo {author} {\bibfnamefont {T.~F.}\ \bibnamefont {Krauss}},\ }\bibfield  {title} {\bibinfo {title} {Charged and neutral exciton complexes in individual self-assembled $\mathrm{In}(\mathrm{Ga})\mathrm{As}$ quantum dots},\ }\href {https://doi.org/10.1103/PhysRevB.63.073307} {\bibfield  {journal} {\bibinfo  {journal} {Phys. Rev. B}\ }\textbf {\bibinfo {volume} {63}},\ \bibinfo {pages} {073307} (\bibinfo {year} {2001})}\BibitemShut {NoStop}%
\bibitem [{\citenamefont {Ediger}\ \emph {et~al.}(2007)\citenamefont {Ediger}, \citenamefont {Bester}, \citenamefont {Gerardot}, \citenamefont {Badolato}, \citenamefont {Petroff}, \citenamefont {Karrai}, \citenamefont {Zunger},\ and\ \citenamefont {Warburton}}]{EdigerWarburton2007}%
  \BibitemOpen
  \bibfield  {author} {\bibinfo {author} {\bibfnamefont {M.}~\bibnamefont {Ediger}}, \bibinfo {author} {\bibfnamefont {G.}~\bibnamefont {Bester}}, \bibinfo {author} {\bibfnamefont {B.~D.}\ \bibnamefont {Gerardot}}, \bibinfo {author} {\bibfnamefont {A.}~\bibnamefont {Badolato}}, \bibinfo {author} {\bibfnamefont {P.~M.}\ \bibnamefont {Petroff}}, \bibinfo {author} {\bibfnamefont {K.}~\bibnamefont {Karrai}}, \bibinfo {author} {\bibfnamefont {A.}~\bibnamefont {Zunger}},\ and\ \bibinfo {author} {\bibfnamefont {R.~J.}\ \bibnamefont {Warburton}},\ }\bibfield  {title} {\bibinfo {title} {Fine structure of negatively and positively charged excitons in semiconductor quantum dots: Electron-hole asymmetry},\ }\href {https://doi.org/10.1103/PhysRevLett.98.036808} {\bibfield  {journal} {\bibinfo  {journal} {Phys. Rev. Lett.}\ }\textbf {\bibinfo {volume} {98}},\ \bibinfo {pages} {036808} (\bibinfo {year} {2007})}\BibitemShut {NoStop}%
\bibitem [{\citenamefont {Warming}\ \emph {et~al.}(2009)\citenamefont {Warming}, \citenamefont {Siebert}, \citenamefont {Schliwa}, \citenamefont {Stock}, \citenamefont {Zimmermann},\ and\ \citenamefont {Bimberg}}]{Bimberg2009}%
  \BibitemOpen
  \bibfield  {author} {\bibinfo {author} {\bibfnamefont {T.}~\bibnamefont {Warming}}, \bibinfo {author} {\bibfnamefont {E.}~\bibnamefont {Siebert}}, \bibinfo {author} {\bibfnamefont {A.}~\bibnamefont {Schliwa}}, \bibinfo {author} {\bibfnamefont {E.}~\bibnamefont {Stock}}, \bibinfo {author} {\bibfnamefont {R.}~\bibnamefont {Zimmermann}},\ and\ \bibinfo {author} {\bibfnamefont {D.}~\bibnamefont {Bimberg}},\ }\bibfield  {title} {\bibinfo {title} {Hole-hole and electron-hole exchange interactions in single inas/gaas quantum dots},\ }\href {https://doi.org/10.1103/PhysRevB.79.125316} {\bibfield  {journal} {\bibinfo  {journal} {Phys. Rev. B}\ }\textbf {\bibinfo {volume} {79}},\ \bibinfo {pages} {125316} (\bibinfo {year} {2009})}\BibitemShut {NoStop}%
\end{thebibliography}%
\end{document}